\documentclass[prd,preprint,tightenlines,floatfix,showpacs,
preprintnumbers,nofootinbib,eqsecnum]{revtex4-1}
\RequirePackage{lineno}

\usepackage[T1]{fontenc}		

\usepackage{amsmath,amsfonts,amssymb,amstext,mathrsfs}
\usepackage{mathpazo}

\usepackage[dvips]{graphicx}
\usepackage{epsf,float}
\usepackage{revsymb}
\usepackage[title]{appendix}
\usepackage{dcolumn}
\usepackage{braket}
\usepackage{color,xcolor}
\usepackage{graphicx}
\usepackage{subfigure}
\usepackage{multirow}
\usepackage{tabularx}
\usepackage{pstricks}
\usepackage[section]{placeins}
\usepackage{booktabs}
\usepackage{array}
\usepackage{todonotes}

\usepackage{hyperref}

\usepackage{pbox}

\newcommand{\bba}{\mbox{\boldmath $b_{1}$}}
\newcommand{\bbb}{\mbox{\boldmath $b_{2}$}}
\newcommand{\bbi}{\mbox{\boldmath $b_{i}$}}
\newcommand{\bb}{\mbox{\boldmath $b$}}

\renewcommand\slash[1]{\not \! #1}

\begin{document}


\title{Anomalous electromagnetic moments of $\tau$ lepton \\
in $\gamma \gamma \to \tau^+ \tau^-$ reaction in Pb+Pb collisions at the LHC}

\author{Mateusz Dynda{\l}}
\email{Mateusz.Dyndal@cern.ch}
\affiliation{CERN, Geneva, Switzerland}

\author{Mariola K{\l}usek-Gawenda}
\email{Mariola.Klusek@ifj.edu.pl}
\affiliation{Institute of Nuclear Physics Polish Academy of Sciences, PL-31342 Krakow, Poland}

\author{Matthias Schott}
\email{mschott@cern.ch}
\affiliation{Johannes Gutenberg University, Mainz, Germany}

\author{Antoni Szczurek\footnote{Also at Faculty of Mathematics and Natural Sciences, University of Rzeszow, ul. Pigonia 1, 35-310 Rzesz\'ow, Poland.}}
\email{Antoni.Szczurek@ifj.edu.pl}
\affiliation{Institute of Nuclear Physics Polish Academy of Sciences, PL-31342 Krakow, Poland}

\date{\today}

\vspace{2cm}

\begin{abstract}
We discuss the sensitivity of the $\gamma \gamma \to \tau^+ \tau^-$  process in ultraperipheral Pb+Pb collisions on the anomalous magnetic ($a_{\tau}$) and electric ($d_{\tau}$) moments
of $\tau$ lepton at LHC energies. 
We derive the corresponding cross sections by folding the elementary cross section with the heavy-ion photon fluxes and considering semi-leptonic decays of both $\tau$ leptons in the fiducial volume of ATLAS and CMS detectors. 
We present predictions for total and differential cross sections, and for the ratios to $\gamma \gamma \to e^+ e^- (\mu^+ \mu^-$) process.
These ratios allow to cancel theoretical and experimental uncertainties when performing precision measurements 
at the LHC.
The expected limits on $a_{\tau}$ with existing Pb+Pb dataset are found to be better by a factor of two comparing to current best experimental limits and can be further improved by another factor of two at High Luminosity LHC.

\end{abstract}

\maketitle

\section{Introduction}

Ultraperipheral collisions (UPC) of heavy-ions provide a very clean environment to study various two-photon induced processes~\cite{Baur:2001jj, Baltz:2007kq}.
Most recent examples include the production of electron pairs~\cite{Adams:2004rz, Afanasiev:2009hy, Abbas:2013oua, Adam:2019mby}, muon pairs~\cite{ATLAS-CONF-2016-025} and light-by-light scattering~\cite{Aaboud:2017bwk, Sirunyan:2018fhl, Aad:2019ock}.
These reactions can also give rise to the production of tau lepton pairs, which provides a highly interesting opportunity to study the electromagnetic properties of the $\tau$ lepton via the Pb+Pb$\rightarrow$Pb+Pb+$\tau^+\tau^-$  process using data from the Large Hadron Collider (LHC).

The presence of $\gamma\tau\tau$ vertex in this reaction gives sensitivity to the anomalous electromagnetic couplings of the tau lepton. 
Since the $\gamma\gamma\rightarrow\tau^+\tau^-$ subprocess diagram contains two such vertices, this reaction provides even an enhanced sensitivity to the anomalous magnetic ($a_{\tau}$) and electric ($d_{\tau}$) moments of the $\tau$ lepton.

The strongest experimental constraints on $a_{\tau}$ come from the kinematics of the similar production process, $e^+e^-\rightarrow e^+e^-\tau^+\tau^-$, measured by the DELPHI collaboration at the LEP2 collider~\cite{Abdallah:2003xd, Cornet:1995pw} yielding a limit of
\begin{equation}
    -0.052 < a_{\tau} < 0.013 ~(95\%~CL)~.
\end{equation}
The experimental limits on $a_{\tau}$ were also derived by the L3 and OPAL collaborations in radiative $Z\rightarrow \tau^+\tau^-\gamma$ events at LEP~\cite{Acciarri:1998iv, Ackerstaff:1998mt}, but they are typically weaker by a factor of two comparing to the DELPHI limits.
For comparison, the theoretical Standard Model (SM) value of $a_{\tau}$ is~\cite{Passera:2006gc, Eidelman:2007sb}:
\begin{equation}
   a_{\tau}^{\textrm{th}} = 0.00117721 \pm 0.00000005~,
\end{equation}
i.e. significantly smaller than the currently available experimental bounds. 

Measuring $a_{\tau}$ with improved precision tests the $\tau$ lepton compositeness~\cite{Silverman:1982ft} and can be sensitive to physics beyond the Standard Model (BSM), including supersymmetric scenarios~\cite{Martin:2001st}, TeV-scale leptoquarks~\cite{Feruglio:2018fxo}, left-right symmetric models~\cite{GutierrezRodriguez:2004ch} and unparticle physics~\cite{Moyotl:2012zz}. 

In the $\gamma\tau\tau$ coupling, another interesting contribution is the CP-violating effects which create electric dipole moment, $d_{\tau}$.
The $d_{\tau}$ arises only at three-loop in SM and is therefore highly suppressed: $|d_{\tau}^{\textrm{th}}|< 10^{-34}$ $e\cdot\textrm{cm}$~\cite{Hoogeveen:1990cb}.
However, various BSM sources of CP violation can enhance $d_{\tau}$~\cite{Appelquist:2004mn, Zhao:2014vga, Dekens:2018bci}.
The presence of electric dipole moment of $\tau$ can be investigated
via studying so-called CP-odd observables in $e^+e^- \to \tau^+\tau^-$ reaction~\cite{Bernreuther:1989kc,Bernreuther:1993nd} and the most stringent limits on $d_{\tau}$ were set by Belle~\cite{Inami:2002ah}.

There are many existing proposals how to improve the experimental sensitivity on $a_{\tau}$ and $d_{\tau}$ using lepton beams and future datasets of Belle-II~\cite{Fael:2013ij, Eidelman:2016aih, Chen:2018cxt}, CLIC~\cite{Koksal:2018env, Gutierrez-Rodriguez:2019umw}, and LHeC~\cite{Gutierrez-Rodriguez:2019umw, Koksal:2018xyi}.
The LHC feasibility studies focus on the usage of proton--proton collisions ~\cite{Samuel:1992fm, Atag:2010ja, Hayreter:2013vna, Galon:2016ngp, Fu:2019utm, Gutierrez-Rodriguez:2019umw}, but also on heavy-ion UPC~\cite{delAguila:1991rm, Beresford:2019gww}.

In this article we study the sensitivity of the $\gamma \gamma \to \tau^+ \tau^-$ process on $a_{\tau}$ and $d_{\tau}$ in Pb+Pb collisions at the LHC.
We present calculations of the cross sections for the nuclear reaction, including outgoing $\tau$ decays and explicit dependence on $a_{\tau}$, and considering fiducial volumes of the ATLAS~\cite{Aad:2008zzm} and CMS~\cite{Chatrchyan:2008aa} detectors. While the authors of~\cite{Beresford:2019gww} rely on an Effective Field Theory (EFT) approach to perform predictions on the sensitivity of LHC UPC data on $a_{\tau}$, we provide a first independent calculation for the Pb+Pb$\rightarrow$Pb+Pb+$\tau^+\tau^-$ process for different $a_{\tau}$ values. We also discuss the strategy to suppress the impact of systematic uncertainties by exploiting the ratio to the $\gamma \gamma \to e^+ e^- (\mu^+ \mu^-$) processes.

\section{Theoretical framework}
\label{sec:theory}

The calculation of the process Pb+Pb$\rightarrow$Pb+Pb+$\tau^+\tau^-$ requires the convolution of the two-photon luminosity with the elementary $\gamma \gamma \rightarrow \tau^+\tau^-$ cross section. 
In our study, the nuclear cross section for the production of $\ell^+\ell^-$ pair 
in ultraperipheral heavy ion collision is calculated 
in the impact parameter space \cite{KlusekGawenda:2010kx} and expressed via the formula:
\begin{eqnarray}
\sigma \left( AA \to AA \ell^+\ell^- ; \sqrt{s_{AA}} \right)&=& 
\int \sigma \left(\gamma\gamma \to \ell^+\ell^-; W_{\gamma\gamma} \right)  
N(\omega_1, \bba) N(\omega_2, \bbb) 
S^2_{abs}(\bb) \nonumber \\
&& \times  \frac{W_{\gamma\gamma}}{2}
dW_{\gamma \gamma} \, d{\rm Y}_{\ell\ell} \, d\overline{b}_x \, d\overline{b}_y 
\, d^2b \,.
\label{eq:sig_nucl_tot}
\end{eqnarray} 
Here $b$ denotes the impact parameter, i.e. the distance between colliding nuclei
in the plane perpendicular to their direction of motion.
$W_{\gamma\gamma}=\sqrt{4\omega_1\omega_2}$ is the invariant
mass of the $\gamma\gamma$ system and $\omega_i$, $i=1,2$,
is the energy of the photon which is emitted from the first or second nucleus, respectively.
${\rm Y}_{\ell\ell}$
is the rapidity of the $\ell^+\ell^-$ system. 
The quantities 
$\overline{b}_x = (b_{1x}+b_{2x})/2$, $\overline{b}_y=(b_{1y}+b_{2y})/2$
are given in terms of $b_{ix}$ and $b_{iy}$ which are the components of 
the $\bba$ and $\bbb$ vectors 
that mark the point (distance from the first and second nucleus) 
where photons collide and particles are produced. 
The absorption factor, $S^2_{abs}(\bb) = \theta(\bb-2R_{Pb})$, assures that only peripheral collisions are considered.
The diagram illustrating these quantities in the impact parameter
space can be found in \cite{KlusekGawenda:2010kx}, where also  the dependence of the photon flux   
$N\left( \omega_i, \bbi \right)$ on the charge form factor is presented.

Our main calculations rely on the \textit{realistic form factor}, defined as the Fourier transform of the charge distribution in the nucleus.
However, more sophisticated calculations are required for differential cross section predictions as well as for predictions in certain fiducial volumes which are typically imposed by experimental constrains. For those predictions, we introduce an additional kinematic parameter related to angular distribution for the subprocess into the underlying integration. A more detailed discussion of the nuclear cross section for the Pb+Pb$\rightarrow$Pb+Pb+$\ell^+\ell^-$ reaction that includes kinematic variables of outgoing leptons is given in~\cite{vanHameren:2017krz}.

In general, we study the $\gamma \gamma \to \ell^+\ell^-$ subprocess 
where the momenta of the incoming photons are denoted by $p_1$ and $p_2$, while $p_3$ and $p_4$ denote the positively and negatively charged lepton momenta, respectively. In addition, we define $p_t=p_2 - p_4 = p_3 - p_1$ and $p_u = p_1 - p_4 = p_3 - p_2$.
The amplitude for the $\gamma \gamma \to \ell^+\ell^-$ reaction in the $t$- and $u$-channel was previously derived \cite{Klusek-Gawenda:2017lgt} and is given by the formula:
\begin{eqnarray}
{\mathcal M}
&=&
(-i)\,
\epsilon_{1 \mu}
\epsilon_{2 \nu}
\,\bar{u}(p_{3}) 
\Big(
i\Gamma^{(\gamma \ell\ell)\,\mu}(p_{3},p_{t})
\frac{i(\slash{p}_{t} + m_{\ell})}{t - m_{\ell}^2+i\epsilon}
i\Gamma^{(\gamma \ell\ell)\,\nu}(p_{t'}-p_{4}) \nonumber\\
&&
+
i\Gamma^{(\gamma \ell\ell)\,\nu}(p_{3},p_{u})
\frac{i(\slash{p}_{u} + m_{\ell})}{u - m_{\ell}^2+i\epsilon}
i\Gamma^{(\gamma \ell\ell)\,\mu}(p_{u'}-p_{4}) \Big)
v(p_{4}) \,.
\label{eq:amp_2to2}
\end{eqnarray}
Here a photon-lepton vertex function is introduced that depends on the momentum transfer, $q=p'-p$. Denoting $p'$ and $p$ as momenta of incoming and outgoing lepton, respectively, this can be written as:
\begin{equation}
i\Gamma^{(\gamma \ell\ell)}_{\mu}(p',p) = 
-ie\left[ \gamma_{\mu} F_{1}(q^{2})+ \frac{i}{2 m_{\ell}}
\sigma_{\mu \nu} q^{\nu} F_{2}(q^{2}) + \frac{1}{2 m_{\ell}} \gamma^5
\sigma_{\mu \nu} q^{\nu}  F_{3}(q^{2})
 \right]
\,,
\label{eq:gamma_lepton_vertex}
\end{equation}
where $\sigma_{\mu \nu} =\dfrac{i}{2}[\gamma_{\mu},\gamma_{\nu}]$ is the spin tensor 
that is proportional to the commutator of the gamma matrices, $F_1(q^2)$ and $F_2(q^2)$
are the Dirac and Pauli form factors, $F_3(q^2)$ is the electric dipole form factor.  
The last term violates CP symmetry and its non-zero value can be evidence of physics BSM.
The asymptotic values of the form factors, in the $q^2 \rightarrow 0$ limit, are the moments describing the electromagnetic properties of the lepton: $F_1(0) = 1$, $F_2(0) = a_{\ell}$ and $F_3(0) = d_{\ell} \frac{2m_\ell}{e}$.
Since the virtualities of exchanged photons for ultraperipheral Pb+Pb collisions at the LHC are very small (typically $Q^2_{1,2}<0.001$~GeV$^2$), this asymptotic condition is well fulfilled.

Finally, the differential elementary cross section for the dilepton production in the $\gamma\gamma$-fusion reaction
is given as follows:
\begin{equation}
\frac{\mathrm{d}\sigma(\gamma \gamma \to \ell^+\ell^-)}{\mathrm{d} z} = \frac{2\pi}{64\pi^2s} \frac{\left| \mbox{\boldmath{$p_{out}$}} \right|}{\left| \mbox{\boldmath{$p_{in}$}} \right|}
\frac{1}{4} \sum_{\rm{spin}} \left| \mathcal{M} \right|^2 \, ,
\end{equation}
where $z=\cos \theta$ and $\theta$ is an angle of the outgoing leptons relative to the beam direction in the photon--photon center-of-mass frame,
$s$ is the invariant mass squared of the $\gamma\gamma$ system,
{\boldmath{$p_{out}$}} and {\boldmath{$p_{in}$}} are the 3-momenta of outgoing (lepton) and initial particle (photon), respectively. 

\begin{figure}[!b]
\centering
\includegraphics[width=0.465\linewidth]{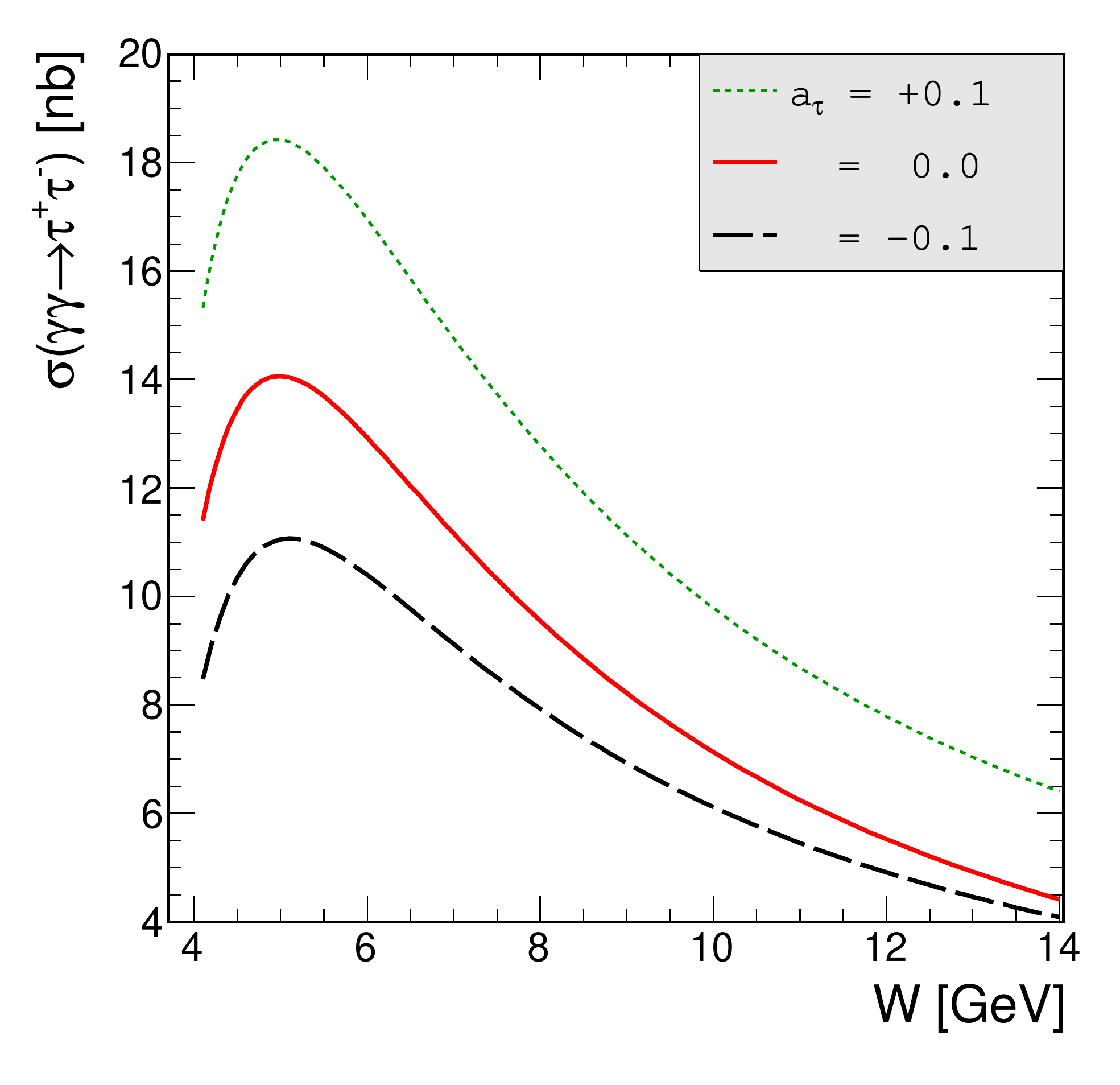}
\includegraphics[width=0.465\linewidth]{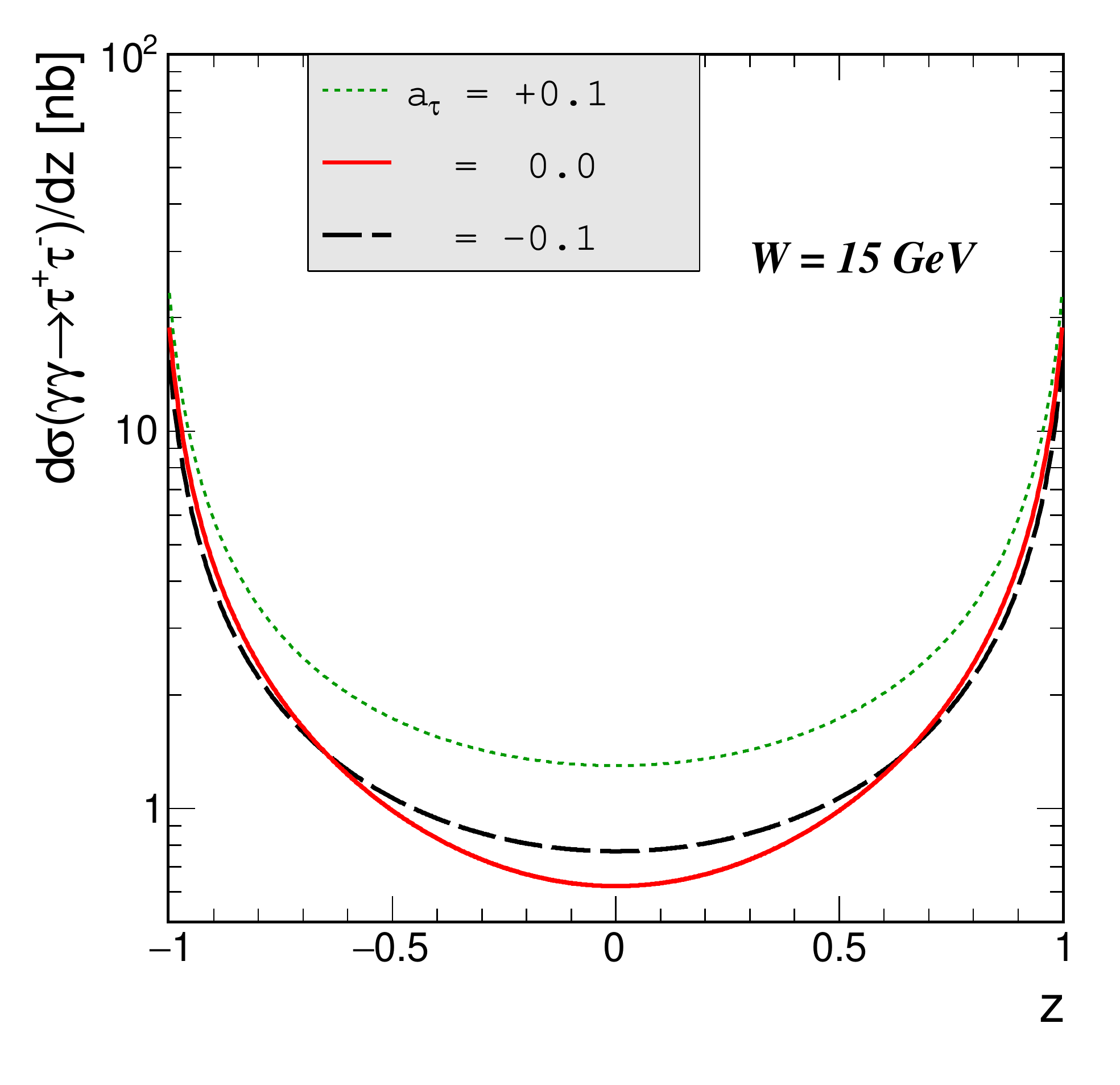}
\caption{Elementary cross section for $\gamma \gamma \to \tau^+\tau^-$ process as a function of $W_{\gamma\gamma}=m_{\tau\tau}$ (left) 
	and as a function of $z = \cos \theta$ for $W_{\gamma\gamma} = 15$~GeV (right).}
\label{fig:elementary_dsig_dz_W15GeV}
\end{figure}

The elementary $\gamma \gamma \to \tau^+\tau^-$ cross section strongly depends on the value of anomalous magnetic moment of $\tau$ lepton.
Figure~\ref{fig:elementary_dsig_dz_W15GeV} illustrates the impact of non-zero $a_{\tau}$ value on
the elementary cross section for $\gamma \gamma \to \tau^+\tau^-$ process as a function of $W_{\gamma\gamma}$ ($=m_{\tau\tau}$) and $\cos \theta$ at $W_{\gamma\gamma} = 15$~GeV.
Shown are the results for three representative values of the anomalous magnetic moment, $a_{\tau} = +0.1$ (green dotted line), $a_{\tau} =0$ (red solid line) and $a_{\tau} =-0.1$ (black dashed line), respectively.

    \label{eq_F123}


\begin{figure}[!b]
\centering
\includegraphics[width=0.5\textwidth]{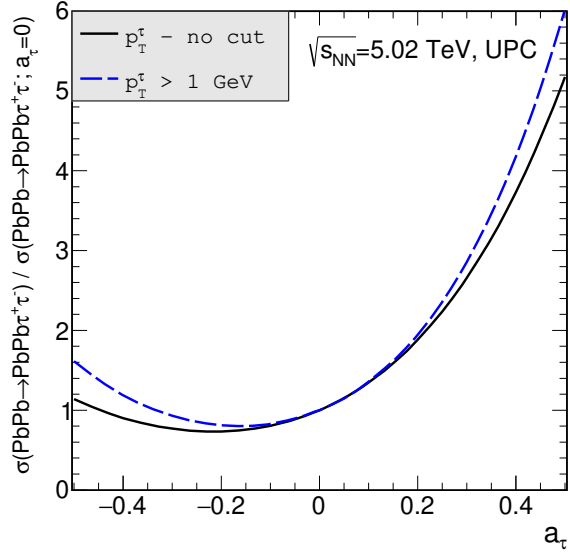}
\caption{Ratio of the total nuclear cross sections for Pb+Pb$\rightarrow$Pb+Pb+$\tau^+\tau^-$ production at the LHC energies as a function of $a_{\tau}$, relative to SM ($a_{\tau}=0$).
The ratio of the cross sections with extra $p_{\mathrm{T}}^{\tau}>1$~GeV requirement applied is also shown.}
\label{fig:ratio_nuclear}
\end{figure}

Figure~\ref{fig:ratio_nuclear} shows the ratio of the total (integrated) nuclear cross section for $\tau$ pair production 
in Pb+Pb collisions at $\sqrt{s_{NN}}=5.02$~TeV with respect to the results with $a_{\tau}=0$ (SM). 
We show the results both for the full momentum space (black solid line)
and with extra requirement of $p_{\textrm{T}}^{\tau}>1$~GeV (blue dashed line). 
The total cross section values for $a_{\tau}=0$ are: $\sigma($Pb+Pb $\to$ Pb+Pb+$\tau^+ \tau^-; p_{\textrm{T}}^{\tau}>0\, \mathrm{\,GeV})=1.06$~mb
and $\sigma($Pb+Pb $\to$ Pb+Pb+ $\tau^+ \tau^-; p_{\textrm{T}}^{\tau}>1\, \mathrm{GeV})=0.73$~mb.
The relative cross section changes significantly with $a_{\tau}$, while its dependence on the $p_{\textrm{T}}^{\tau}$ cut value is relatively small for $\left| a_{\tau} \right|<0.1$.  

\begin{figure}[!b]
\centering
\includegraphics[width=0.52\textwidth]{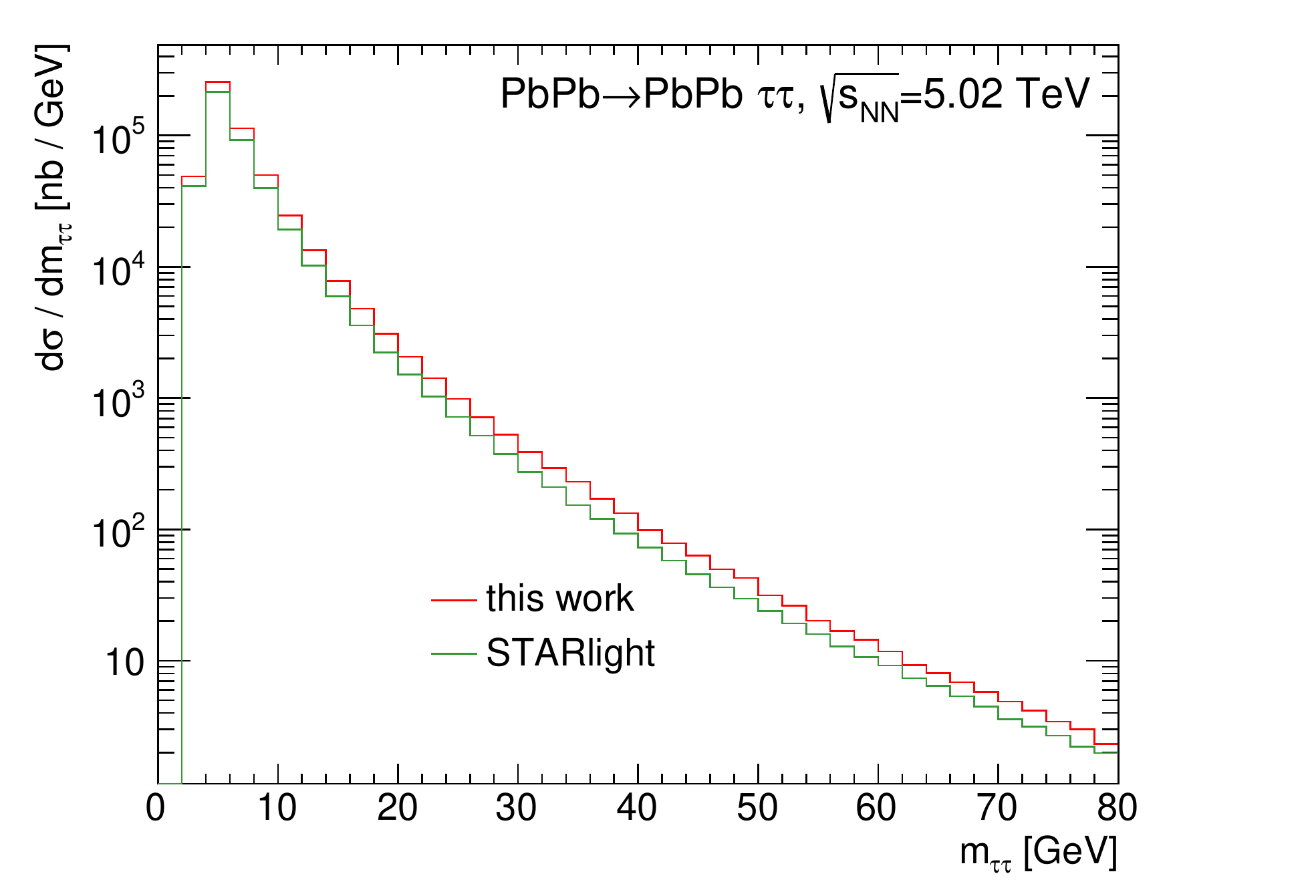}
\hspace{-1cm}
\includegraphics[width=0.52\textwidth]{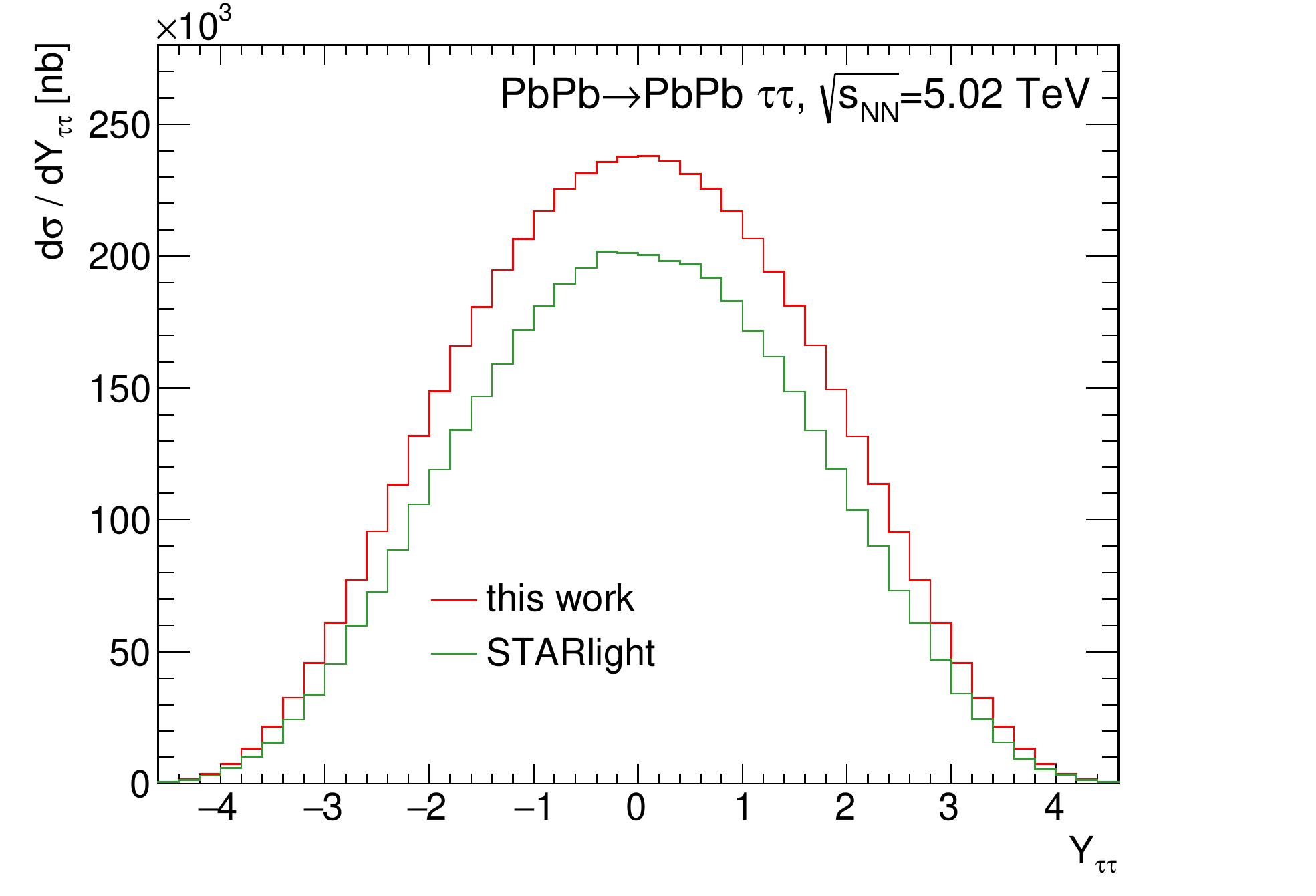}
\caption{Total cross section for Pb+Pb$\rightarrow$Pb+Pb+$\tau^+\tau^-$ production at the LHC energies as a function of ditau invariant mass (left) and ditau rapidity (right). Our results are compared with the results obtained from the \textsc{STARlight} MC generator. 
}
\label{fig:mass_rapidity_taus}
\end{figure}

We also compare our results ($a_{\tau}=0$) with \textsc{STARlight} Monte Carlo (MC) generator~\cite{Klein:2016yzr}, which is commonly used to describe ultraperipheral heavy-ion collision data. 
Figure~\ref{fig:mass_rapidity_taus} shows the comparison of total cross section for Pb+Pb$\rightarrow$Pb+Pb+$\tau^+\tau^-$ production as a function of $m_{\tau\tau}$ and $Y_{\tau\tau}$. 
In general, the predictions from \textsc{STARlight} are systematically lower by about 20\% in comparison to the results of the calculations described above. The overall shape of the $m_{\tau\tau}$ distribution is also slightly different between the two calculations.
This is mainly because \textsc{STARlight} applies extra $|\bba|>R_{Pb}$ and $|\bbb|>R_{Pb}$ requirements in the modelling of photon fluxes, on top of the $S^2_{abs}(\bb)$ requirement (see Eq.~(\ref{eq:sig_nucl_tot})).
However, as it will be shown in Sec.~\ref{sec:results}, the modeling uncertainty of incoming photon fluxes cancel out to a large extent, once the ratio of various $\gamma \gamma \to \ell^+\ell^-$ ($\ell=e,~\mu,~\tau$) cross sections is used.

\section{Fiducial selection and $\tau$ decays}
\label{selection}
In order to study the experimental sensitivity on $a_\tau$ in the $\gamma\gamma\rightarrow\tau^+\tau^-$ processes at the LHC, one has to record UPC events, which contain two reconstructed tau candidates and no further activity in the detector. Since the tau is the heaviest lepton with a lifetime of $3 \times 10^{-13}$ s, it decays into lighter leptons (electron or muon) or hadrons (mainly pions and kaons) before any direct interaction with the detector material. The reconstruction of tau candidates depends therefore on the identification of its unique decay signatures. The primary $\tau$ decay channels produce one charged particle in the final state, denoted as $1ch$, or one-prong decays in the following,
\begin{equation}
    \tau^\pm\rightarrow \nu_{\tau} +\ell^\pm+ \nu_{\ell} ~(\ell=e,~\mu)~,
    \end{equation}
    \begin{equation}
    \tau^\pm\rightarrow \nu_{\tau}+ \pi^{\pm} + n\pi^{0}~,
\end{equation}
or three charged particles, denoted as $3ch$, or three-prong decays, i.e.
\begin{equation}
    \tau^\pm\rightarrow \nu_{\tau}+ \pi^{\pm}+\pi^{\mp}+\pi^{\pm} + n\pi^{0} \, .
\end{equation}
Approximately 80\% of all $\tau$ decays are the one-prong decays and 20\% of them are the three-prong decays. 

While the differential cross sections of Pb+Pb$\rightarrow$Pb+Pb+$\tau^+\tau^-$ at the LHC are based on the previous calculations, the \textsc{Pythia8.243} program~\cite{Sjostrand:2014zea} is used to model $\tau$ decays for our studies, as it simulates all known $\tau$ decay channels with a branching fraction greater than 0.04\%, including large number of 2- to 6-body decay modes~\cite{Ilten:2012zb}. The effect of QED final state radiation (FSR) from outgoing leptons is also simulated by \textsc{Pythia8}. The effect of spin correlations for $\tau$ decays is not taken into account, as this feature is currently not supported for the $\gamma\gamma\rightarrow\tau^+\tau^-$ process within \textsc{Pythia8} framework.
However, an attempt to estimate the size of this effect is made and is discussed in details in Appendix~\ref{appendixa}.

\begin{figure}[ht!]
\centering
\includegraphics[width=0.49\textwidth]{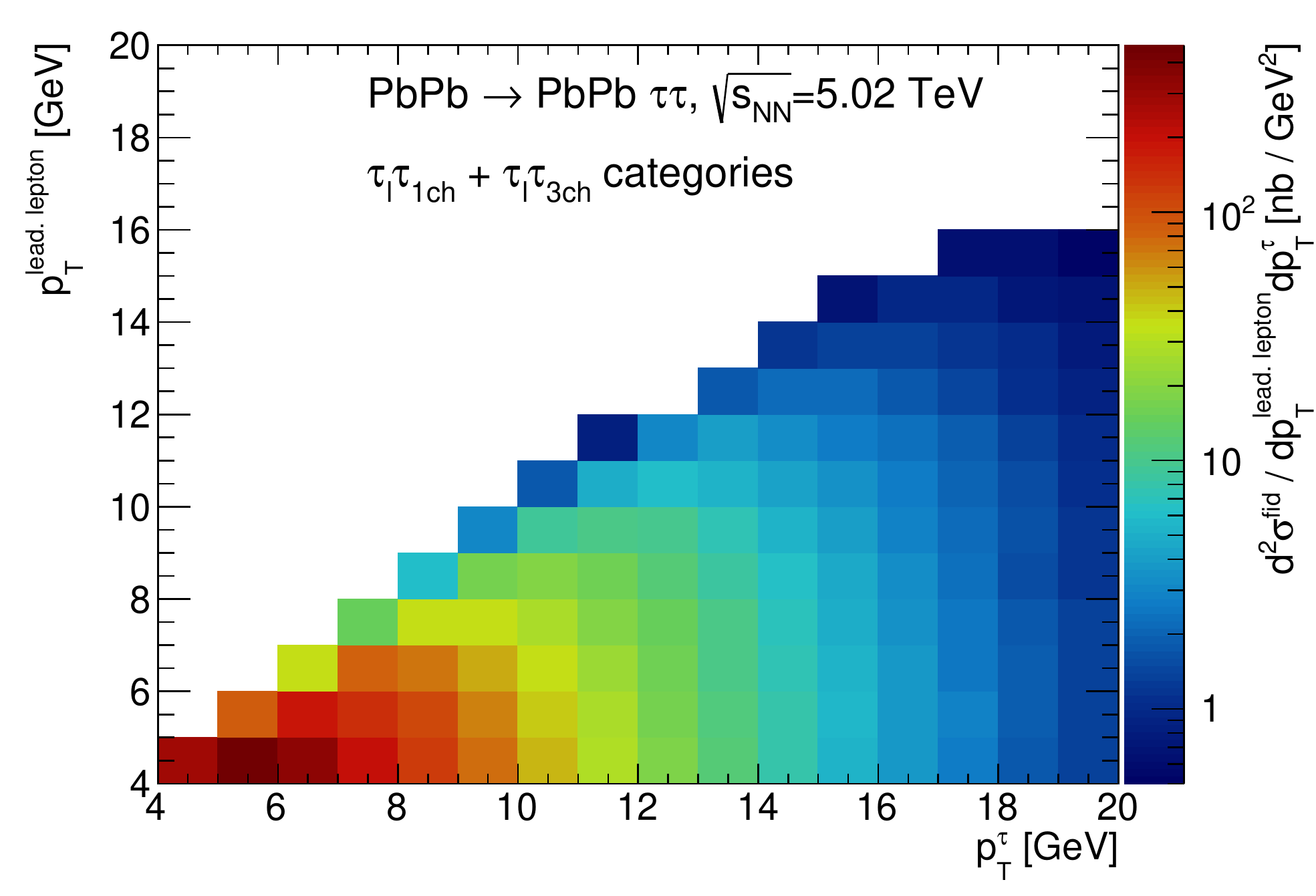}
\includegraphics[width=0.49\textwidth]{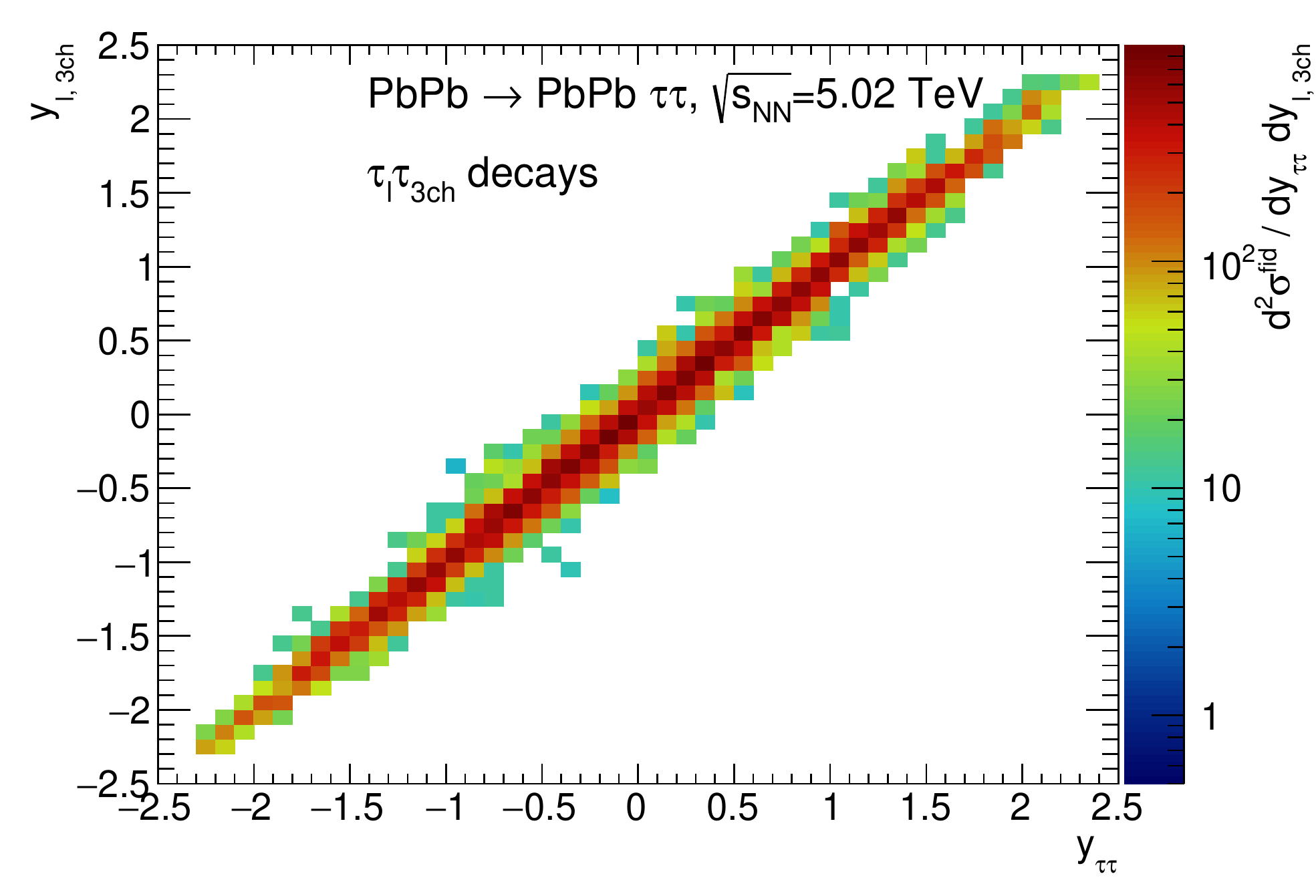}
\caption{(left) Correlation between $p_{\mathrm{T}}$ of the $\tau$ lepton and  $p_{\mathrm{T}}$ of the electron (muon) from its decay for both event categories.
(right) Correlation between the rapidity of $\tau^+\tau^-$ system and  the rapidity of the final-state charged-particle system for $\tau_{\ell}\tau_{3ch}$ category.
The results are obtained for SM scenario ($a_{\tau}=0$) with full set of fiducial cuts applied.}
\label{fig:taupt_lpt_corr}
\end{figure}

We propose that the $\gamma\gamma\rightarrow\tau^+\tau^-$ candidates events are selected by requiring at least one $\tau$ lepton to decay leptonically, as this allows that existing triggering algorithms of the ATLAS or CMS detector can be used~\cite{Aad:2008zzm, Chatrchyan:2008aa}. 
The leading electron or muon is further required to have a transverse momentum of $p_{\mathrm{T}}>4$~GeV and $|\eta|<2.5$ to allow for an efficient reconstruction and identification by the LHC detectors. 
The correlation between the $p_{\mathrm{T}}$ of one tau and its charged decay lepton is shown in Fig.~\ref{fig:taupt_lpt_corr}, indicating a broad smearing of the decay lepton $p_{\mathrm{T}}$ due to the presence of neutrinos. 
On the contrary, there is a good correlation between the rapidity of $\tau^+\tau^-$ system and the rapidity of the final-state charged-particle system, as shown in Fig.~\ref{fig:taupt_lpt_corr} (right). 

It should be noted that the majority of produced $\tau$ lepton pairs have relatively low energy/transverse momentum. Therefore, the standard $\tau$ identification tools, developed by the ATLAS and CMS collaborations~\cite{Aad:2015unr, Sirunyan:2018pgf} are not expected to be applicable. We propose therefore to categorize the $\gamma\gamma\rightarrow\tau^+\tau^-$ candidate events by their decay mode: $\tau_{\ell}\tau_{1ch}$\footnote{due to the presence of soft lepton tracks, experimentally these leptons cannot be easily distinguished from pions, hence fully leptonic decays of both $\tau$ are kept as a part of a more generic $\tau_{\ell}\tau_{1ch}$ category} or $\tau_{\ell}\tau_{3ch}$. All charged-particle tracks from $\tau_{1ch}$ or $\tau_{3ch}$ decays are required to have a transverse momentum of $p_{\mathrm{T}}>0.2$~GeV and a pseudo-rapidity of $|\eta|<2.5$.

Possible background processes which could fake the  $\gamma\gamma\rightarrow\tau^+\tau^-$signal are: the two-photon quark-antiquark production ($\gamma\gamma\rightarrow q\bar{q}$) and the (semi)exclusive production of electron/muon pairs. As demonstrated already in Ref.~\cite{Beresford:2019gww}, the $\gamma\gamma\rightarrow q\bar{q}$ have a significantly larger charged-particle multiplicity than the signal and hence this background is fully reducible by exclusivity requirements.

\begin{figure}[ht!]
\centering
\includegraphics[width=0.52\textwidth]{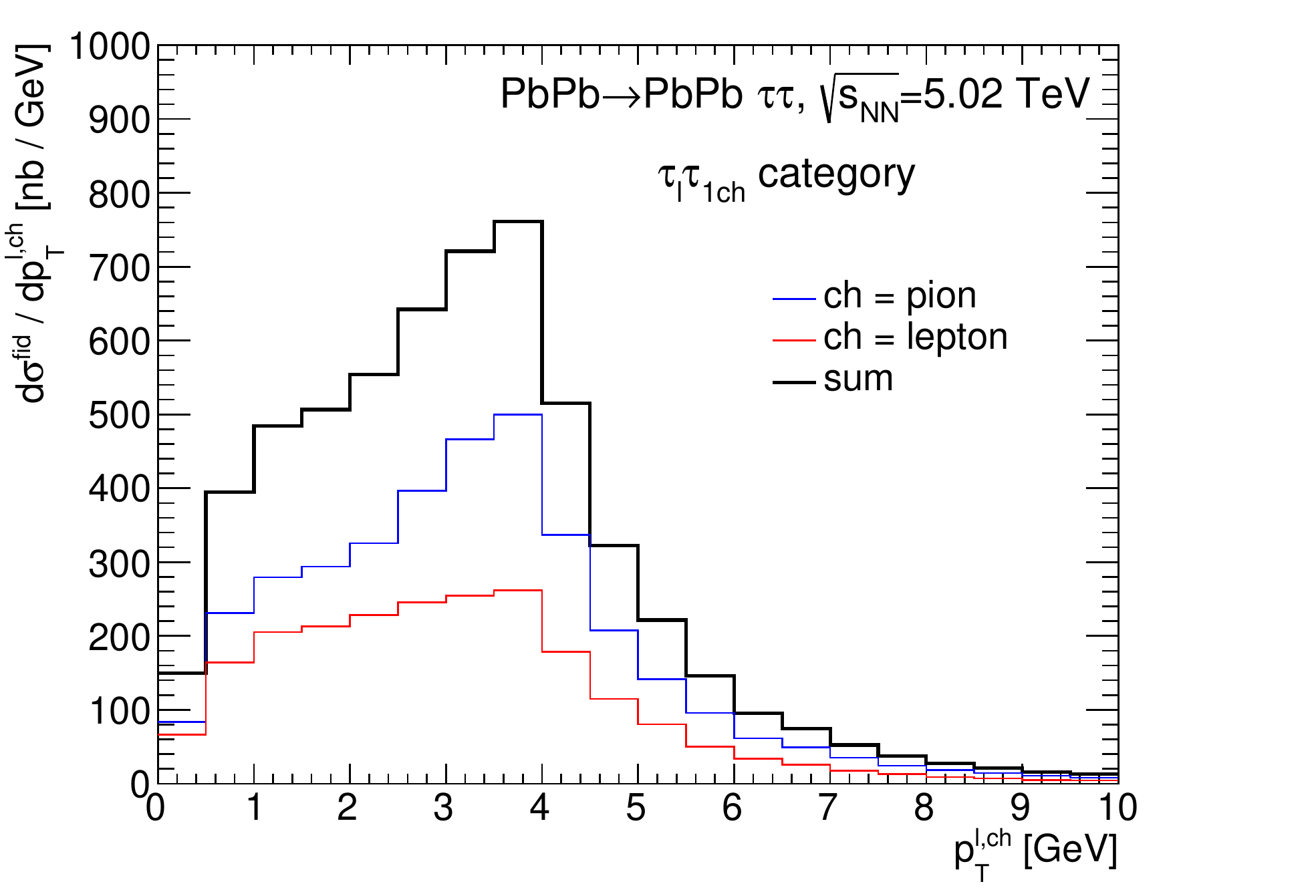}
\hspace{-1cm}
\includegraphics[width=0.52\textwidth]{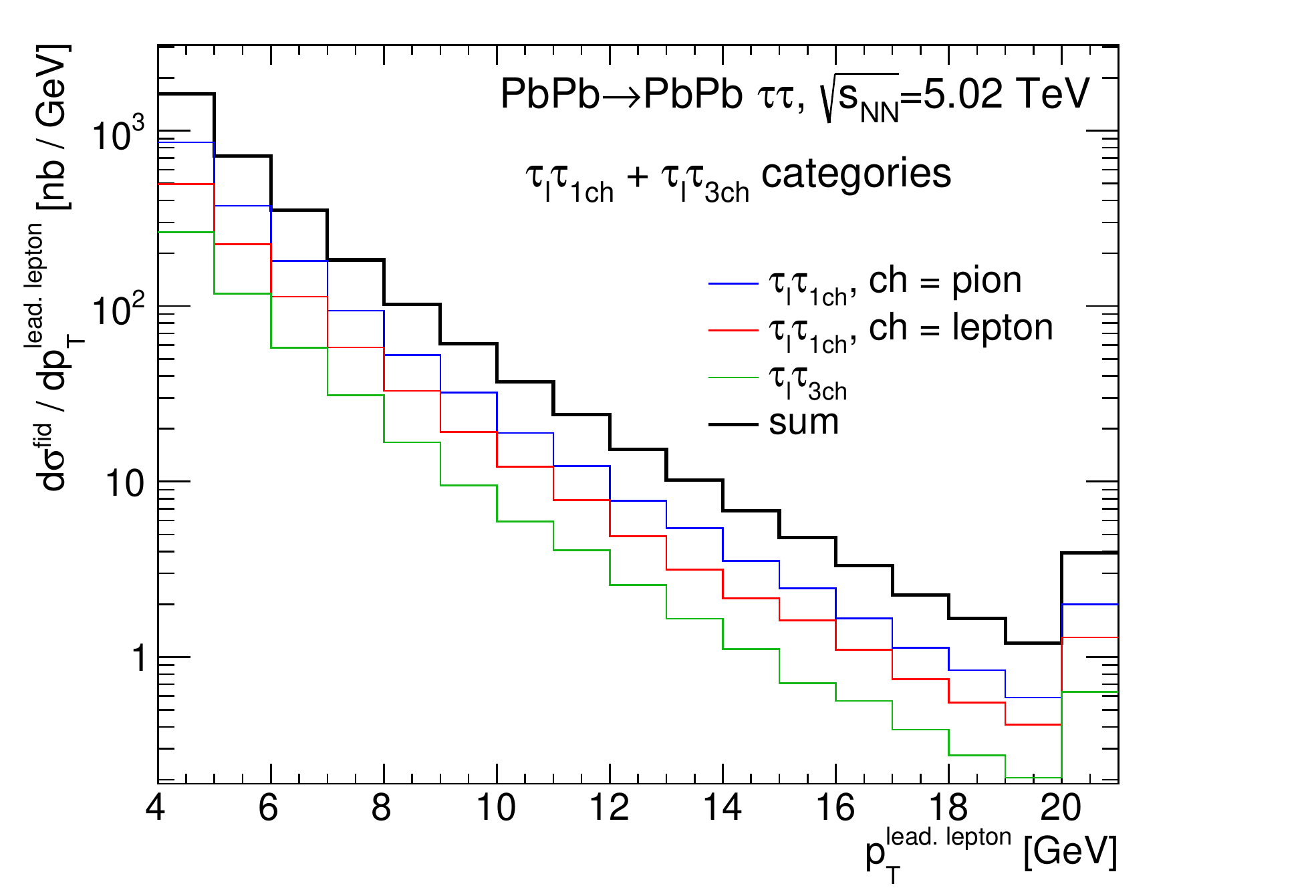}
\caption{(Left) Fiducial cross section as a function of $p_{\mathrm{T}}$ of the lepton+track system ($p_{\mathrm{T}}^{\ell~ch}$) in the $\tau_{\ell}\tau_{1ch}$ category for SM scenario ($a_{\tau}=0$) and before applying $p_{\mathrm{T}}^{\ell~ch}>1$~GeV requirement. (Right) Fiducial cross section as a function of $p_{\mathrm{T}}$ of the leading lepton for various event categories and the SM scenario ($a_{\tau}=0$). The last bin denotes integrated fiducial cross section above $p_{\mathrm{T}}^{lead~lepton}=20$~GeV.}
\label{fig:ltrk_pt}
\end{figure}

On the other hand, the $\gamma\gamma\rightarrow\ell^+\ell^-$ production can become an irreducible background for the $\tau_{\ell}\tau_{1ch}$ category. To suppress this background, additional requirements on $p_{\mathrm{T}}$ of the lepton+track system ($p_{\mathrm{T}}^{\ell~ch}>1$~GeV) have to be applied for this event category.
As presented in Fig.~\ref{fig:ltrk_pt}, an increased $p_{\mathrm{T}}^{\ell~ch}>1$~GeV cut removes only 10\% of signal events, however, suppresses at the same time significantly back-to-back $\gamma\gamma\rightarrow e^+e^-~(\mu^+\mu^-)$ processes and leading therefore to negligible background contribution from this process.

A further possible source of background is the semi-coherent dilepton production, i.e. $\gamma^{*}\gamma\rightarrow\ell^+\ell^-$. Here, the $p_{\mathrm{T}}$ of the dilepton system can be as large as $p_{\mathrm{T}}^{\ell~ch}$ in the signal process. Due to relatively large momentum transfer from $\gamma^{*}$, the outgoing ion dissociates, emitting forward neutrons detectable in ZDC systems~\cite{ATLAS:2007aa}. Thus a requirement of zero neutrons in both ion directions provides a straightforward way to estimate and even fully suppress this background.
Since the neutrons can be occasionally emitted also in the signal process due to extra Coulomb exchanges~\cite{Baltz:2009jk}, a full neutron veto can lead to 20--30\% reduction of the cross sections.
The study of the neutron emissions is, however, beyond the scope of this paper.

The differential fiducial cross section as a function of leading lepton $p_{\mathrm{T}}$ for all event categories is also shown in Fig.~\ref{fig:ltrk_pt}. 
The majority of selected events have a single charged pion in the final state due to the relatively large branching fraction. These events are followed by fully leptonic decays of both $\tau$ leptons. 
The $\tau_{\ell}\tau_{1ch}$ category has the lowest cross sections. The integrated fiducial cross section values are 2630 nb for $\tau_{\ell}\tau_{1ch}$ category (including 1650 nb for $\ell^\pm\pi^{\mp}$ decays, 980 nb for $\ell^\pm\ell^\mp$ decays) and 515 nb for $\tau_{\ell}\tau_{3ch}$ category, assuming Pb+Pb collision energy of $\sqrt{s_{NN}}=5.02$~TeV.

%

\section{Results and discussion}
\label{sec:results}

\begin{figure}[ht!]
\centering
\includegraphics[width=0.52\textwidth]{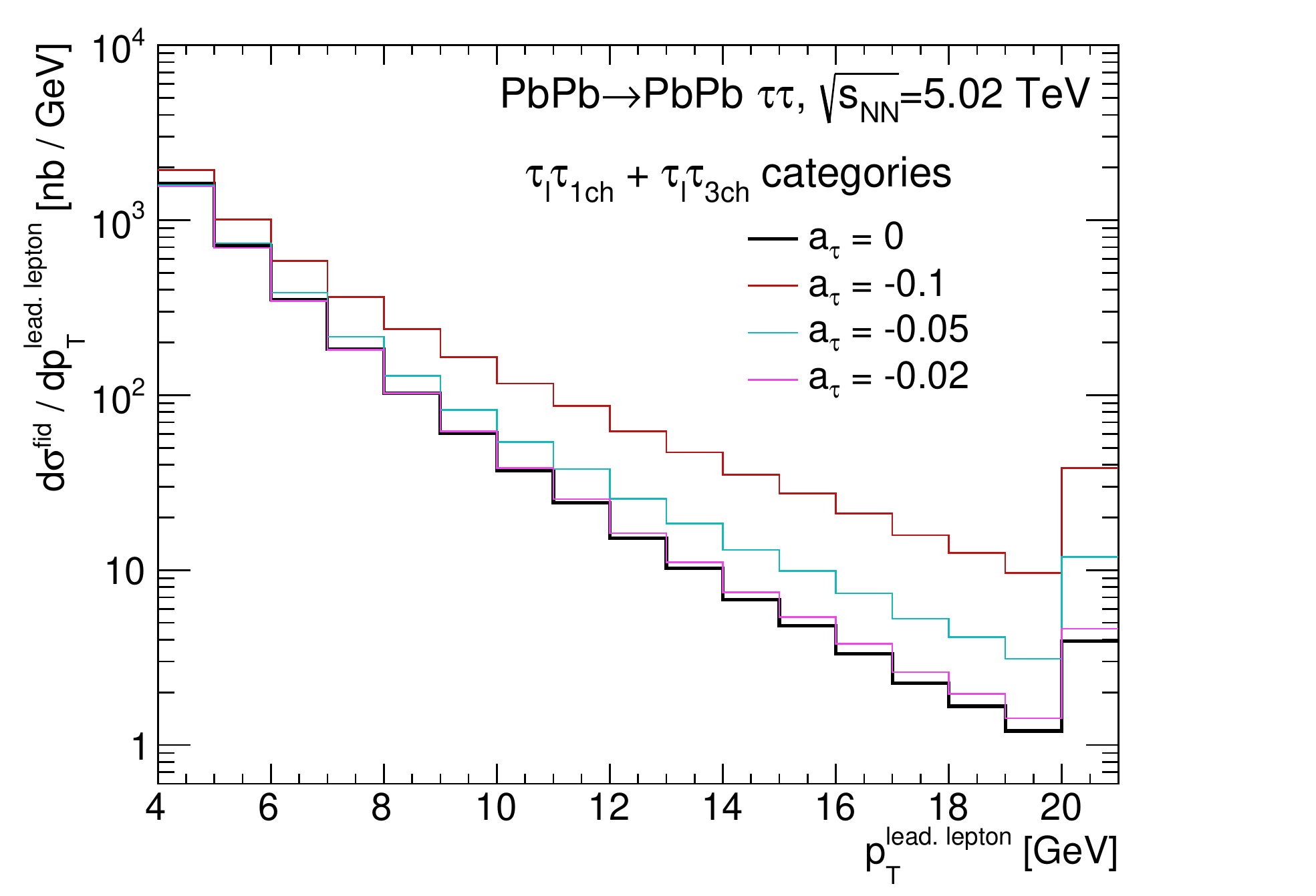}
\hspace{-1cm}
\includegraphics[width=0.52\textwidth]{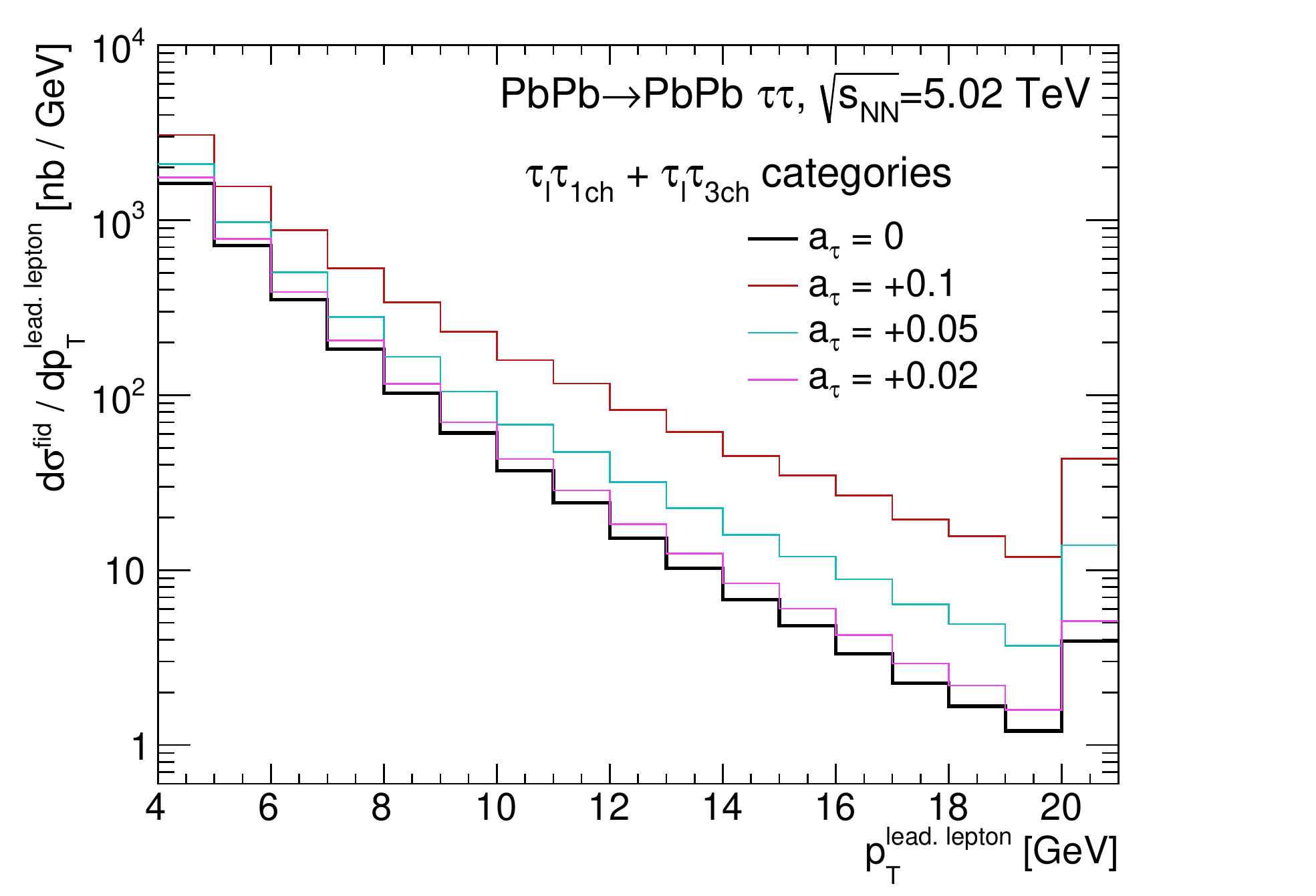}
\caption{Fiducial cross section as a function of $p_{\mathrm{T}}$ of the leading lepton for all event categories summed together and various $a_{\tau}$ values: $a_{\tau}=-0.1,~-0.05,~-0.02,~0$ (left) and $a_{\tau}=0,~0.02,~0.05,~0.1$ (right). The last bin denotes integrated fiducial cross section above $p_{\mathrm{T}}^{lead~lepton}=20$~GeV.}
\label{fig:pt_vs_atau}
\end{figure}

Figure~\ref{fig:pt_vs_atau} summarizes the fiducial cross section for Pb+Pb$\rightarrow$Pb+Pb+$\tau^+\tau^-$ process as a function of the leading lepton $p_{\mathrm{T}}$ ($p_{\mathrm{T}}^{lead~lepton}$) for SM scenario ($a_{\tau}=0$) as well as other representative values of $a_{\tau}$ ($a_{\tau} = -0.1,~-0.05,~-0.02,~0.02,~0.05,~0.1$).
In addition to the overall cross section enhancement, induced by non-zero $a_{\tau}$, there is an interesting change in the shape of $p_{\mathrm{T}}^{lead~lepton}$ distribution visible.
This is due to the fact that the anomalous $\tau$ couplings are more sensitive to higher $\tau$ energies, based on the term $\sigma^{\mu\nu}q_{\nu}$ in Eq.~(\ref{eq:gamma_lepton_vertex}).
There is also an asymmetry between the cross sections for positive and negative $a_{\tau}$ values, which is due to 
interference between the SM part and the anomalous $\tau$ coupling (see Fig.~\ref{fig:ratio_nuclear}).

The integrated fiducial cross sections for different $a_{\tau}$ values are summarized in Table~\ref{tab:numbers}. This table also lists the expected number of reconstructed events in ATLAS or CMS, assuming 80\% reconstruction efficiency within the fiducial region and two values of integrated luminosity ($L_{int}$): 
$L_{int}=2$~nb$^{-1}$ (existing LHC Pb+Pb dataset) or $L_{int}=20$~nb$^{-1}$ (expected High Luminosity LHC, HL-LHC, dataset) at $\sqrt{s_{NN}}=5.02$~TeV.
With the existing Pb+Pb dataset we expect each experiment to reconstruct about 5000 $\gamma\gamma\rightarrow\tau^+\tau^-$ events ($a_{\tau}=0$). The expected number of reconstructed $\tau$ pairs grows to about 50\,000 at the HL-LHC.

\begin{table}[t]
\begin{center}
\begin{tabular}{|l|c|c|c|}
\hline
$a_{\tau}$ value & $\sigma_{fid}$ [nb] & \pbox{20cm}{Expected events \\ ($L_{int}=2$~nb$^{-1}$, $C=0.8$)} & \pbox{20cm}{Expected events \\ ($L_{int}=20$~nb$^{-1}$, $C=0.8$)} \\
\hline
$-0.1$ &   4770 & 7650 & 76\,500 \\
$-0.05$ &  3330 & 5350 & 53\,500 \\
$-0.02$ &  3060 & 4900 & 49\,000 \\
\hline
$0$ (SM) &  3145 & 5050 & 50\,500 \\
\hline
$+0.02$ &  3445 & 5500 & 55\,000 \\
$+0.05$ & 4350 & 6950 & 69\,500 \\
$+0.1$ & 7225 & 11550 & 115\,500 \\
\hline
\end{tabular}
\end{center}
\caption{Integrated fiducial cross sections for Pb+Pb$\rightarrow$Pb+Pb+$\tau^+\tau^-$ process for different $a_{\tau}$ values. The expected number of events assuming 80\% selection efficiency and $L_{int}=2$~nb$^{-1}$ (current LHC Pb+Pb dataset) or $L_{int}=20$~nb$^{-1}$ (expected HL-LHC dataset) are also shown.}
\label{tab:numbers}
\end{table}

\begin{figure}[b!]
\centering
\includegraphics[width=0.52\textwidth]{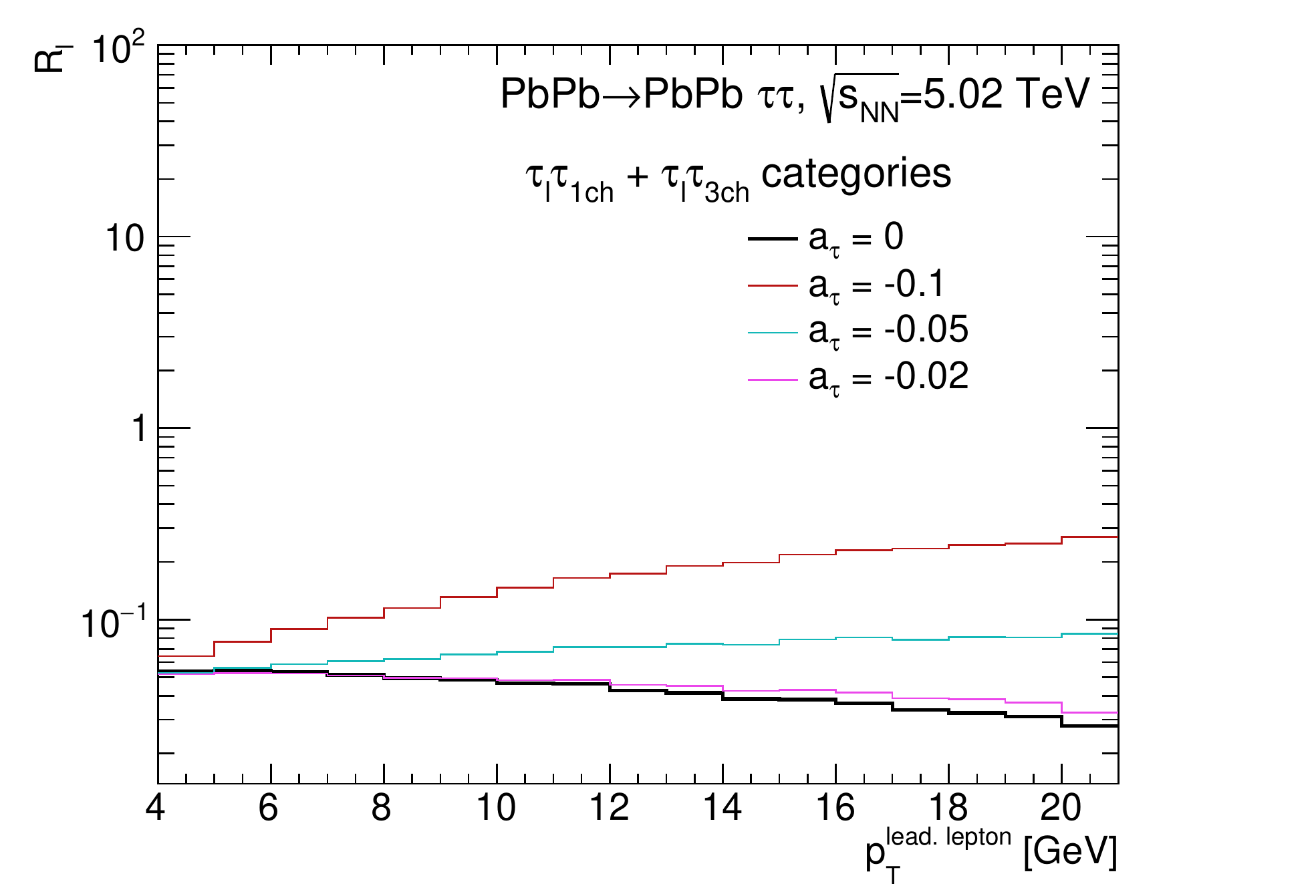}
\hspace{-1cm}
\includegraphics[width=0.52\textwidth]{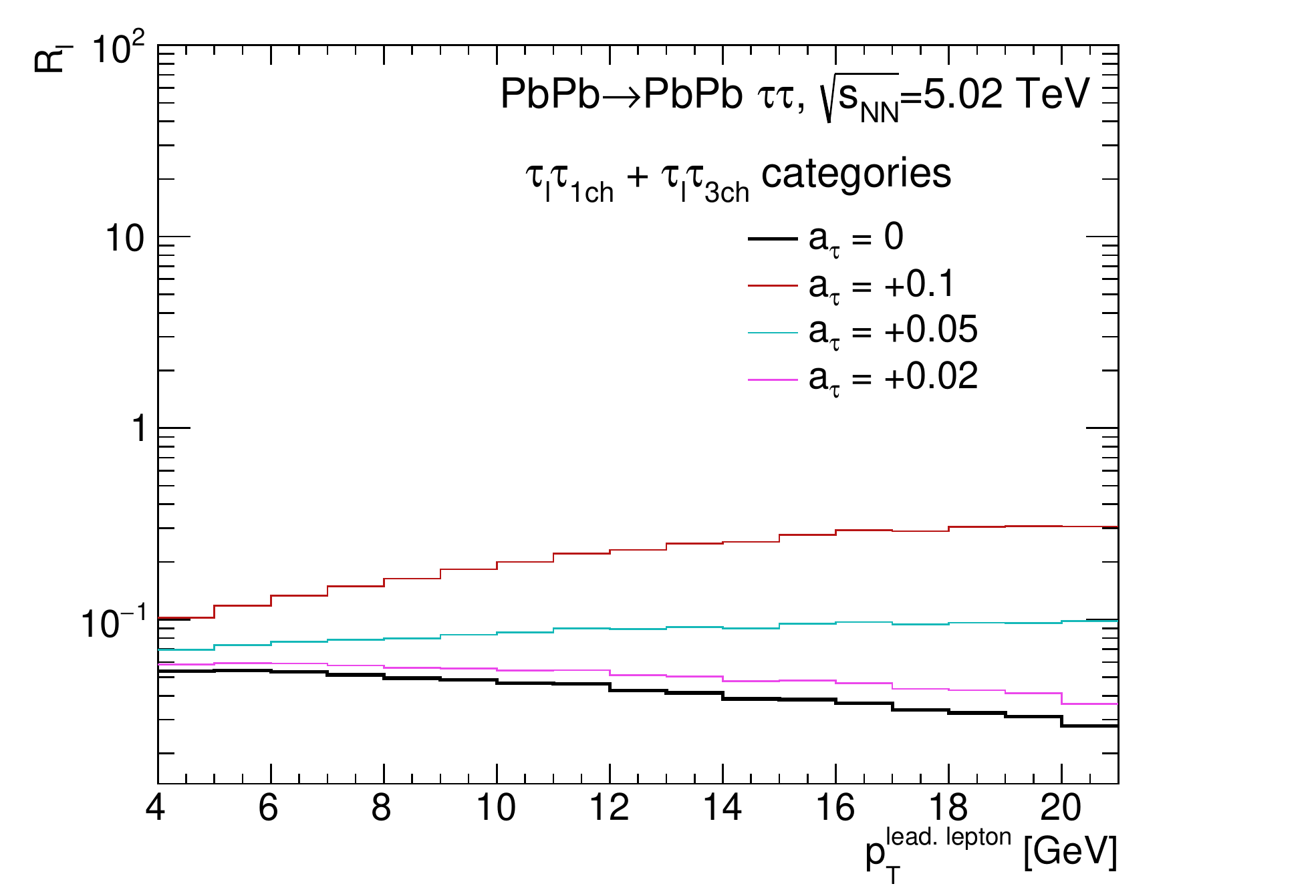}
\caption{Ratio of the fiducial cross sections between $\gamma\gamma\rightarrow\tau^+\tau^-$ and $\gamma\gamma\rightarrow\ell^+\ell^-$ ($\ell=e$ or $\mu$) processes as a function of $p_{\mathrm{T}}$ of the leading lepton for all event categories summed together and different $a_{\tau}$ values: $a_{\tau}=-0.1,~-0.05,~-0.02,~0$ (left) and $a_{\tau}=0,~0.02,~0.05,~0.1$ (right). The last bin denotes the ratio of integrated fiducial cross sections above $p_{\mathrm{T}}^{lead~lepton}=20$~GeV.}
\label{fig:pt_vs_atau_ratio}
\end{figure}

Figure~\ref{fig:pt_vs_atau_ratio} shows the ratio (denoted as $R_{\ell}$) of fiducial cross sections presented in Fig.~\ref{fig:pt_vs_atau} to the fiducial cross sections from the standard candle process $\gamma\gamma\rightarrow\ell^+\ell^-$ ($\ell=e$ or $\mu$).
To calculate these cross sections, the same theoretical framework described in Sec.~\ref{sec:theory} is used. The QED FSR effect is modeled by \textsc{Pythia8}. To match the fiducial selection of $\gamma\gamma\rightarrow\tau^+\tau^-$ signal process, each lepton is required to have $p_{\mathrm{T}}>4$~GeV and $|\eta|<2.5$.
The advantage of studying the cross section ratios is the cancellation of several systematic uncertainties, such as the error on integrated luminosity, uncertainties related to lepton reconstruction, but also theoretical uncertainties, e.g. those that are associated with modeling of the initial photon fluxes.

The fiducial cross sections for $\gamma\gamma\rightarrow e^+ e^-$ and $\gamma\gamma\rightarrow\mu^+\mu^-$ processes are found to be identical, hence one can equally use either of the process.
Experimentally, one can match $\tau_{e}\tau_{1(3)ch}$ channels with $e^+ e^-$ events and $\tau_{\mu}\tau_{1(3)ch}$ channels with $\mu^+ \mu^-$ events to maximize cancellation of systematic uncertainties.

The sensitivity of the $a_{\tau}$ measurement on modeling of initial photon fluxes can be tested by repeating the analysis using the photon--photon luminosity prediction from the \textsc{STARlight}~\cite{Klein:2016yzr} program. As already demonstrated in Sec.~\ref{sec:theory}, the differences in the cross sections between \textsc{STARlight} and the results presented in this work can be as large as 20\%, mainly due to extra requirements applied in the modelling of single photon flux in \textsc{STARlight}.

\begin{figure}[b!]
\centering
\includegraphics[width=0.52\textwidth]{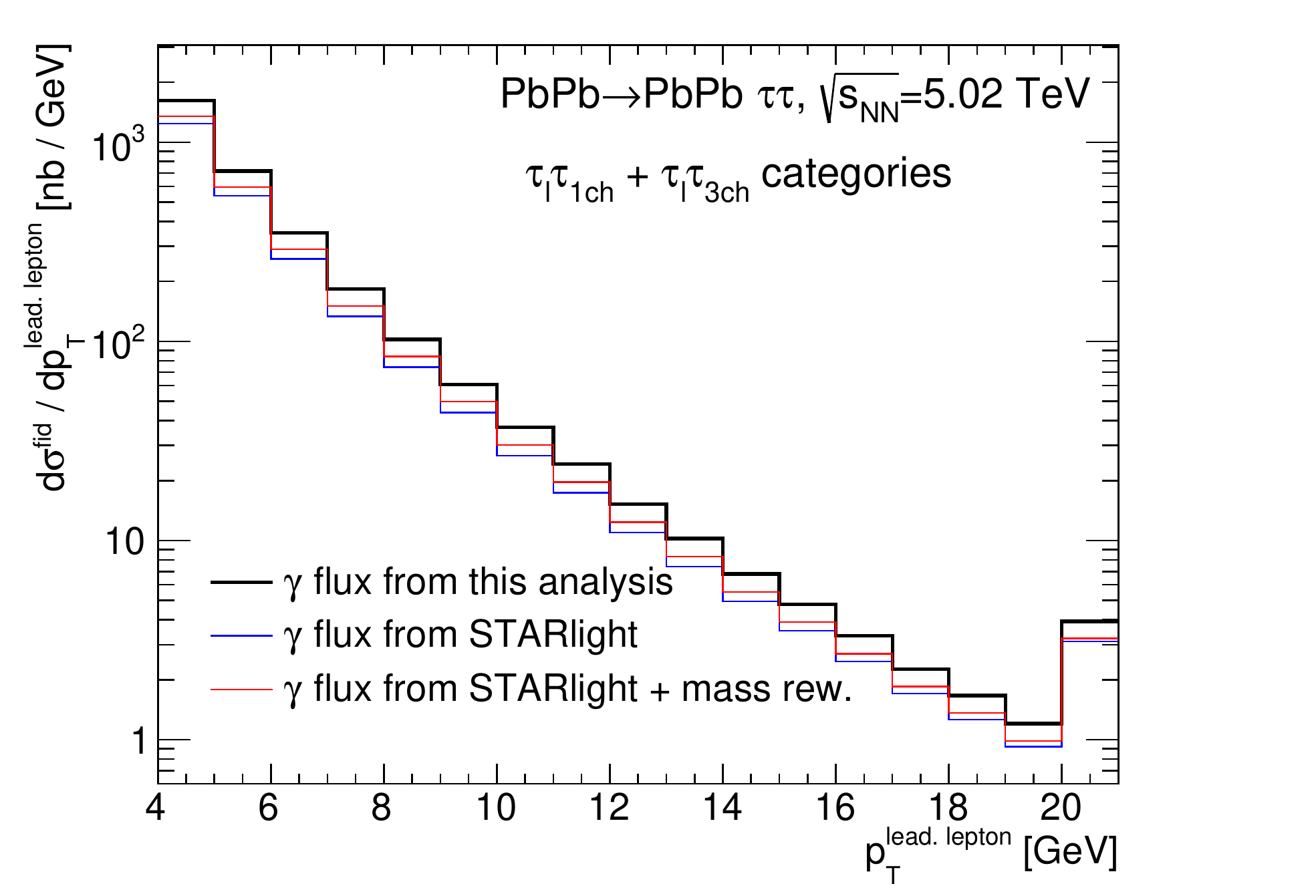}
\hspace{-1cm}
\includegraphics[width=0.52\textwidth]{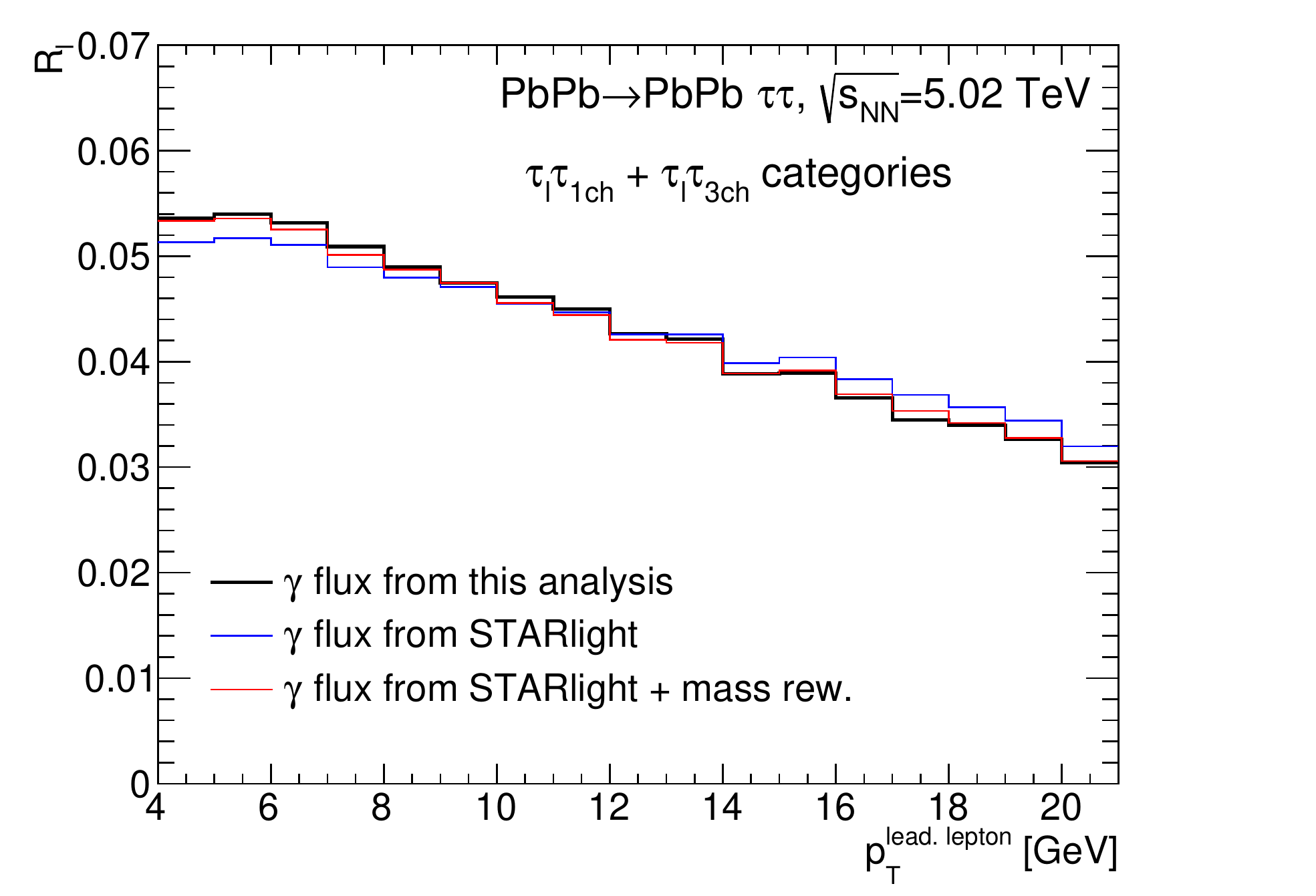}
\caption{Fiducial cross section for the $\gamma\gamma\rightarrow\tau^+\tau^-$ process (left) and the ratio of fiducial cross sections between $\gamma\gamma\rightarrow\tau^+\tau^-$ and $\gamma\gamma\rightarrow\ell^+\ell^-$ ($\ell=e$ or $\mu$) processes (right) as a function of $p_{\mathrm{T}}$ of the leading lepton for all event categories summed together and various photon fluxes. 
The red curve shows the results with extra $m_{\ell\ell}$ shape reweighting as described in the text.
The last bin denotes integrated fiducial cross section (or the ratio) above $p_{\mathrm{T}}^{lead~lepton}=20$~GeV.}
\label{fig:pt_vs_starlight}
\end{figure}

Figure~\ref{fig:pt_vs_starlight} shows the fiducial cross section for $\gamma\gamma\rightarrow\tau^+\tau^-$ process in Pb+Pb UPC at the LHC and its ratio ($R_{\ell}$) to the fiducial cross section from $\gamma\gamma\rightarrow\ell^+\ell^-$ ($\ell=e$ or $\mu$) process for the two choices of initial photon fluxes.
As expected, the difference in the absolute value of the fiducial cross sections is about 20\%. 
However, after taking the ratio to $\gamma\gamma\rightarrow\ell^+\ell^-$  process, the difference becomes suppressed to 5\%.
The remaining difference can be explained by the $m_{\ell\ell}$ shape difference between two implementations (as demonstrated already in Figure~\ref{fig:mass_rapidity_taus}) and the fact that the $p_{\mathrm{T}}$ of the lepton from $\tau$ decay does not necessarily correspond to the $p_{\mathrm{T}}$ of lepton from $\gamma\gamma\rightarrow\ell^+\ell^-$ process.
It is also demonstrated in Figure~\ref{fig:pt_vs_starlight} that an extra reweighting of the shape of $m_{\ell\ell}$ distribution would lead to differences in the ratio that are less than 1\%.
However, it should be noted that in reality the $m_{\ell\ell}$ spectrum can be reweighted directly to the experimental data, thus reducing significantly the impact of theory modelling uncertainties on the measurement.

The expected number of events from Table~\ref{tab:numbers} can be translated into expected sensitivity for probing $a_{\tau}$.
We use the \textsc{RooFit} toolkit~\cite{Verkerke:2003ir} for the statistical analysis of the results.
We perform fits to $R_{\ell}(p_{\mathrm{T}}^{lead~lepton})$ distribution by treating SM results ($a_{\tau}=0$) as background and the difference between $a_{\tau}=0$ and $a_{\tau}=X$ distributions as signal.
A test statistic based on the profile likelihood ratio~\cite{Cowan:2010js} is used under the Asimov approximation.
The procedure exploits both normalization and $p_{\mathrm{T}}^{lead~lepton}$ shape differences, providing extra sensitivity on $a_{\tau}$ measurement.
We use two values of expected systematic uncertainty (5\% and 1\%) and two assumptions on Pb+Pb integrated luminosity (2 nb$^{-1}$ to reflect existing ATLAS/CMS dataset, or 20 nb$^{-1}$ for HL-LHC expectations).

\begin{figure}[htb]
    \centering
    \begin{minipage}{0.49\linewidth}
        \centering
\includegraphics[width=1.0\textwidth]{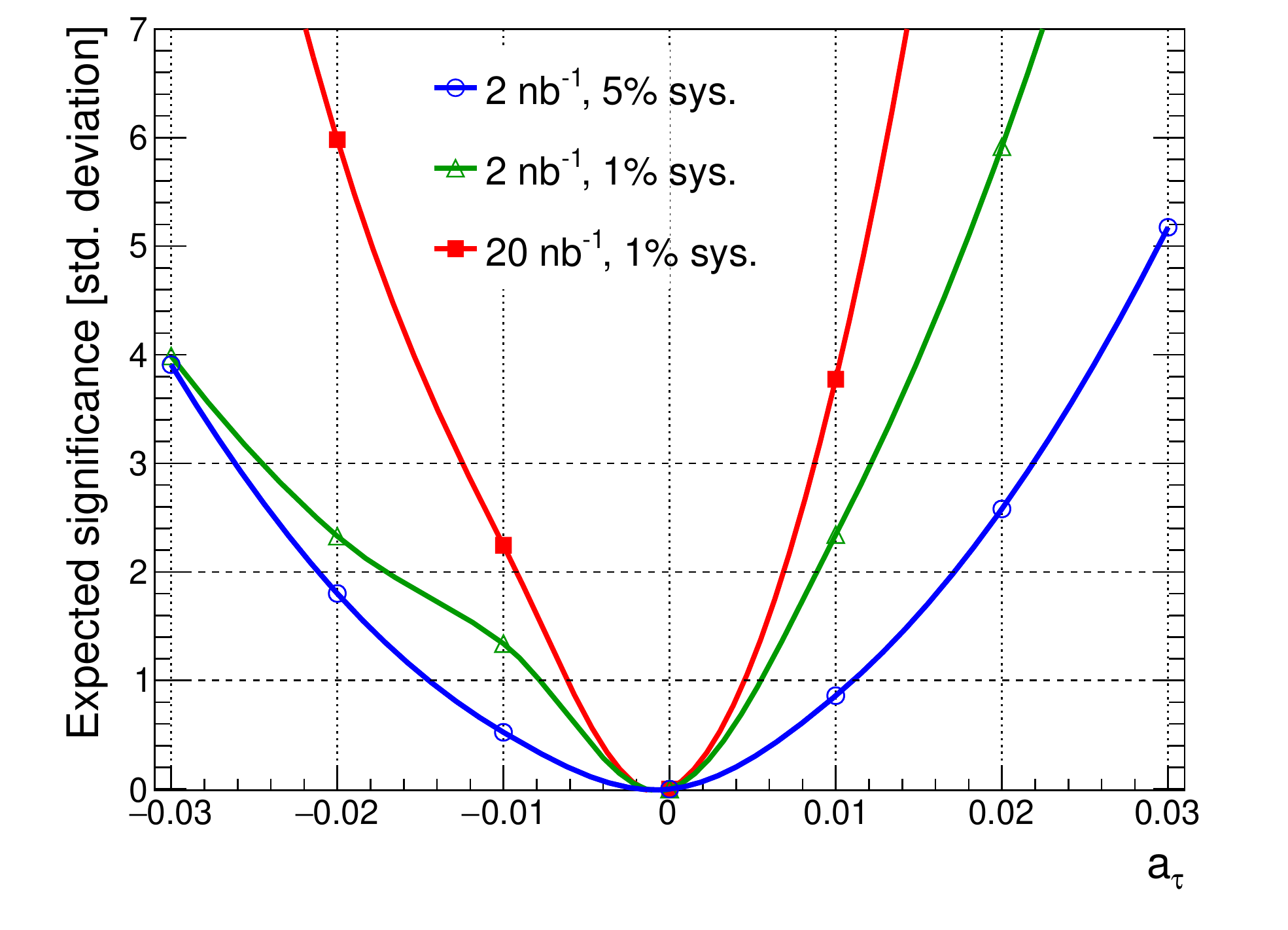}
\caption{\label{fig:sig}Expected signal significance as a function of $a_{\tau}$ for various assumptions on Pb+Pb integrated luminosity (2 nb$^{-1}$ or 20 nb$^{-1}$) and total systematic uncertainty (5\% or 1\%).\color{white} w w w w w w w w w w w w w w w w w w w w w w w w w w w w w w w w w w w w w w w w w w w w w w w}
    \end{minipage}
\hspace{0.1cm}
    \begin{minipage}{0.49\linewidth}
        \centering
\includegraphics[width=1.0\textwidth]{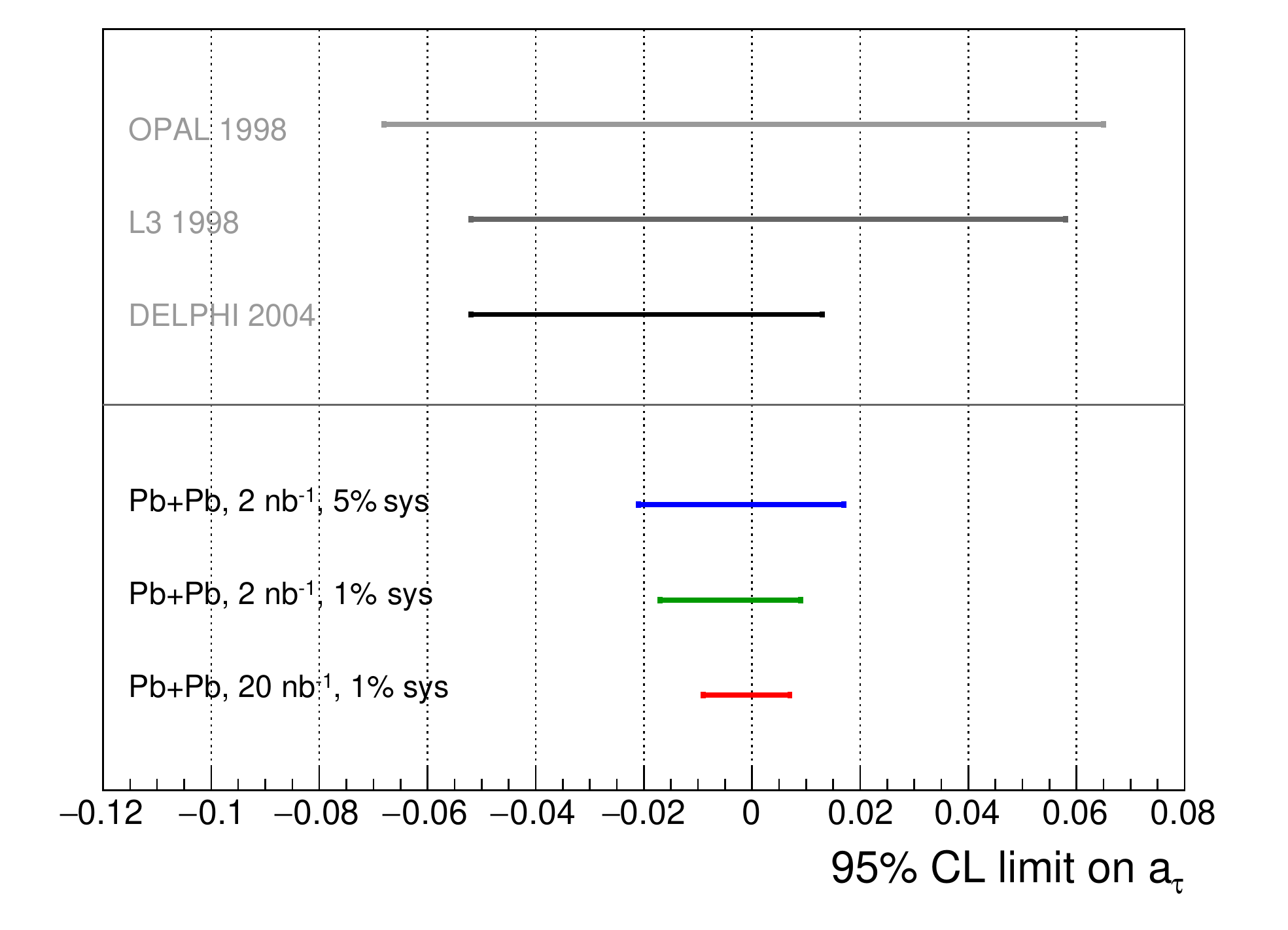}
\caption{\label{fig:limits}Expected 95\% CL limits on $a_{\tau}$ measurement for various assumptions on Pb+Pb integrated luminosity (2 nb$^{-1}$ or 20 nb$^{-1}$) and total systematic uncertainty (5\% or 1\%). Comparison is also made to the existing limits from OPAL~\cite{Ackerstaff:1998mt}, L3~\cite{Acciarri:1998iv} and DELPHI~\cite{Abdallah:2003xd} experiments at LEP.}
    \end{minipage}
\end{figure}

Figure~\ref{fig:sig} shows expected signal significance as a function of $a_{\tau}$. The observed asymmetry in sensitivity between positive and negative $a_{\tau}$ values reflects the destructive interference between SM and the anomalous $\tau$ coupling.

The expected significance can be directly transformed into expected 95\% CL limits on $a_{\tau}$, shown in Fig.~\ref{fig:limits}. Assuming 2 nb$^{-1}$ of integrated Pb+Pb luminosity and 5\% systematic uncertainty, the expected limits are $-0.021<a_{\tau}<0.017$, approximately two times better than DELPHI limits~\cite{Abdallah:2003xd}.
By collecting more data (20 nb$^{-1}$) and with improved systematic uncertainties, these limits can be further improved by another factor of two.
The expected results by studying ultraperipheral collisions at the LHC have therefore the potential to significantly improve the existing limits on $a_{\tau}$.

In addition, using the same methods we study the sensitivity on tau lepton electric dipole moment, $d_{\tau}$.
Our expected 95\% CL sensitivity on $|d_{\tau}|$ assuming $a_{\tau}=0$ is: $|d_{\tau}|<6.3~(4.4)\cdot10^{-17}$ $e\cdot\textrm{cm}$ at the LHC with 5\% (1\%) systematic uncertainty and $|d_{\tau}|<3.5\cdot10^{-17}$ $e\cdot\textrm{cm}$ at HL-LHC (1\% systematic uncertainty). 
For comparison, the current best limits are measured by Belle experiment~\cite{Inami:2002ah}: $-2.2 < Re(d_{\tau}) < 4.5 ~(10^{-17}~e\cdot\textrm{cm})$ and $-2.5 < Im(d_{\tau}) < 0.8 ~(10^{-17}~e\cdot\textrm{cm})$.
Our projected results on $d_{\tau}$ can be therefore competitive with Belle limits.

The expected limits on $a_{\tau}$ and $d_{\tau}$ are found to be approximately factor of two weaker than those reported in Ref.~\cite{Beresford:2019gww}.
This likely points to the issue with EFT approach and the conversion used between the relevant EFT operators and $a_{\tau}$ when calculating elementary $\gamma \gamma \rightarrow \tau^+\tau^-$ cross section.

In the present analysis we have ignored spin-spin correlation effect.
In the Appendix~\ref{appendixa} we have performed
a preliminary study of this effect for fully leptonic decay channel ($\tau^+\tau^-\rightarrow \mu^+ \mu^-$).
We found that spin-spin correlations should
be small for our studies of searches for anomalous magnetic moment of $\tau$ with their rather specific kinematic requirements. 
In general, the spin correlation studies for $\gamma\gamma\rightarrow\tau^+\tau^-$ subprocess are interesting but go beyond the scope
of the present letter and will be performed in details in future as a separate project.

\section{Conclusions}
In this paper, we derived a prediction on the differential cross section of the $\gamma \gamma \to \tau^+ \tau^-$ process and its dependence on anomalous electromagnetic couplings of the tau lepton in ultraperipheral Pb+Pb collisions at the LHC. In contrast to previous calculations, which are based on an Effective Field Theory approach, our calculation is derived from first principles and yields a significantly different inclusive cross section dependence on $a_\tau$ than previously reported \cite{Beresford:2019gww}. 
We also investigated the expected sensitivity on $a_{\tau}$ and $d_{\tau}$, assuming standard LHC detectors using the currently available as well as future datasets. In particular we propose to use cross section ratios of the $\gamma \gamma \to \tau^+ \tau^-$  and $\gamma \gamma \to e^+ e^- (\mu^+\mu^-)$ processes to probe $a_{\tau}$, as several systematic uncertainties cancel and the experimental knowledge of $a_e$ and $a_\mu$ is several orders of magnitude more precise than $a_\tau$ itself. Our studies suggest that the currently available datasets of the LHC experiments are already sufficient to improve the sensitivity on $a_\tau$ by a factor of two, hence, we consider this analysis as highly interesting and worthwhile to be done in the future.
Future Belle-II experiment should give much better constraints on
$|a_{\tau}|<1.75\cdot10^{-5}$ and $|d_{\tau}|<2.04\cdot10^{-19}$~$e\cdot\textrm{cm}$ 
\cite{Chen:2018cxt}.


\section*{Acknowledgements}

We are indebted to Otto Nachtmann for a fruitful discussion.
We are indebted to Anna Kaczmarska for a discussion about
spin correlations and Jakub Zaremba for generation of events with the help 
of the \textsc{Tauola} code.
This study was partially supported by the Polish National Science Center grant
UMO-2018/31/B/ST2/03537 and by the Center for Innovation and Transfer of Natural Sciences and Engineering Knowledge in Rzeszów. 

\begin{appendices}
\section{Study of $\tau^+\tau^-$ spin correlations}
\label{appendixa}

In the present exploratory calculations we consider two
independent (isotropic) decays of $\tau^+$ and $\tau^-$ leptons 
as performed by \textsc{Pythia8}.
Some spin correlations may be of interest in this context.
Such correlations are being considered recently e.g. for the decay
of Higgs boson into $\tau^+ \tau^-$ \cite{BBK2014} where the calculations are much simpler than for the current case. 
The spin correlations were also studied for $e^+ e^- \to \tau^+ \tau^-$ process \cite{BNO93}.
There exists a special computer framework \textsc{TauSpinner} \cite{CPW2012,KPPRS1014}
dedicated to Higgs, $Z$ and $W$ boson decays to tau leptons.
According to our knowledge, the spin--spin correlations were never done for 
the $\gamma \gamma \to \tau^+ \tau^-$ 
(sub)process and no available MC generators have such an option.

The spin correlations of $\tau^+ \tau^-$ decays is an interesting
topic which requires further studies for our $A A \to A A \tau^+ \tau^-$
reaction which, in general, goes beyond the scope of the present letter.
In the present letter we shall discuss (for illustration only) 
the effect of spin-correlations in an approximate way for fully leptonic decays: 
$\tau^+ \to \mu^+ \overline{\nu}_{\tau} \nu_{\mu}$ and
$\tau^- \to \mu^- {\nu}_{\tau} \overline{\nu}_{\mu}$.
For the weak decays discussed here:
\begin{equation}
\frac{d \sigma}{d z} \left( z \right) \ne
\frac{d \sigma}{d z} \left(-z \right) \; ,
\label{not_symmetric}
\end{equation}
where $z = cos(\theta)$ in the $\tau^+$ or $\tau^-$ rest frames with
respect to spin direction
(for illustration see e.g.\cite{MSZB2020}).

To perform decays we do the operational replacement:
\begin{equation}
\hat{\sigma}_{\gamma \gamma \to \tau^+ \tau^-} \rightarrow
\sum_{\lambda_1,\lambda_2}
\hat{\sigma}_{\gamma \gamma \to \tau^+ \tau^-}(\lambda_1,\lambda_2)
\cdot P_{\lambda_1,\lambda_2}(p_{\mu^+/\tau^+}, p_{\mu^-/\tau^-})
\; ,
\end{equation}
where $P_{\lambda_1,\lambda_2}$ are probability densities of the
combined decays. We assume ``independent'' decays, i.e.
\begin{equation}
P_{\lambda_1,\lambda_2}(p_{\mu^+/\tau^+}, p_{\mu^-/\tau^-})
\approx P_{\lambda_1}(p_{\mu^+/\tau^+})
        P_{\lambda_2}(p_{\mu^-/\tau^-})  \; .
\end{equation}
The probability densities are taken from the \textsc{Tauola} code \cite{TAUOLA}:
\begin{eqnarray}
P_{\lambda_1}(p_{\mu^+/\tau^+}) &=& P_{\lambda_1}(p_1^*,z_1^*) \; ,
\nonumber \\
P_{\lambda_2}(p_{\mu^-/\tau^-}) &=& P_{\lambda_2}(p_2^*,z_2^*) 
\end{eqnarray}
and depend on momenta of $\mu^{\pm}$ in the $\tau^{\pm}$ rest frames
($p_k^* = \sqrt{(E_k^*)^2 - m_{\mu}^2}$) and $z_k^* = cos(\theta_k^*)$
with polar angles with respect to $\tau^{\pm}$ spin polarization.
One should note that within the above approximation we neglect azimuthal angle correlations
between two decay planes.

The decays are done with the \textsc{Tauola} MC code \cite{TAUOLA}.
We use the \textsc{Tauola} sample of 10$^5$ events for two different polarizations
of $\tau^+$ and $\tau^-$. 
We calculate matrix elements for a given set of helicities of 
$\tau^+$ and $\tau^-$. Each combination of helicities is treated
separately, i.e. the decays are performed using distributions obtained with \textsc{Tauola} for a given spin polarization.

Our simplified procedure is as follows. 
First we generate weighted events for the $A A \to A A \tau^+ \tau^-$
reaction for a given polarization of $\tau^+$ and $\tau^-$.
The decay is done by the MC method with energy and angular 
distributions in the $\tau$ lepton rest frame taken from the \textsc{Tauola}
program. We are interested only in the muons in the final state.
By performing Lorentz boosts, the momenta of muons in the rest frame of $\tau$ leptons
are transformed to overall center-of-mass system.
As a reference, we also generate a sample for unpolarized $\tau^+ \tau^-$ decays.

We have performed calculations of several single-muon
distributions, as well as many $\mu^+ \mu^-$ correlation observables,
such as:
$d \sigma/d \phi_{\mu^+ \mu^-}$, $d \sigma / d m_{\mu^+ \mu^-}$,
$d \sigma/d y_{\mu^+ \mu^-}$ or $d \sigma/d p_{\textrm{T}}^{\mu^+ \mu^-}$.
In all cases we have observed only up to a few percent effect which is difficult to vizualize. 
For example, the cross section for $a_{\tau}$ = 0 
with neglecting spin correlations and with $p_{\textrm{T}}^{\mu}>$ 4 GeV requirement applied to the leading muon is
7.8288 $BF(\tau \to \mu)^2$ $\mu$b, to be compared with 
7.8350 $BF (\tau \to \mu)^2$ $\mu$b when including spin
correlations as described above.
For the full set of kinematic cuts proposed in this paper (see Section~\ref{selection}), the cross section is
6.773 $BF(\tau \to \mu)^2$ $\mu$b and 6.769 $BF(\tau \to \mu)^2$ $\mu$b,
respectively.
In this case (very specific cuts), we find the effect of spin correlations to be very small.

The potential (neglected) effect of spin correlations in the azimuthal angle between decay planes should vanish for our azimuthal-symmetric cuts. 
Its inclusion for our reaction with many decay channels included 
would be a significant technical effort.

\end{appendices}
\bibliographystyle{unsrturl}
\bibliography{biblio}

\end{document}